\documentclass[%
 reprint,
 superscriptaddress,
nofootinbib,
 amsmath,amssymb,
 aps,
 prl,
floatfix,
twocolumn,
notitlepage
]{revtex4-2}
\usepackage{bbm}
\usepackage{lipsum}
\usepackage{graphicx}
\usepackage{dcolumn}
\usepackage{subfigure}
\usepackage{bm}
\usepackage{breakurl}
\usepackage{enumitem}
\usepackage[normalem]{ulem}
\usepackage{multirow}
\usepackage[multidot]{grffile}

\usepackage{hyperref}
\usepackage{geometry}
\newcommand{\be}{\begin{equation}}
\newcommand{\ee}{\end{equation}}

\def\bc{\begin{center}}
\def\ec{\end{center}}
\def\bea{\begin{eqnarray}}
\def\eea{\end{eqnarray}}

\newcommand{\abs}[1]{\left|#1\right|}

\newcommand{\expec}[1]{\langle #1\rangle}

\def\multiset#1#2{\ensuremath{\left(\kern-.3em\left(\genfrac{}{}{0pt}{}{#1}{#2}\right)\kern-.3em\right)}}

\newcommand{\bigbinom}[2]{
  \Biggl( \!
  \begin{array}{c} 
    #1 \\ #2 
  \end{array} \!
  \Biggr)
}

\usepackage{dcolumn}
\usepackage[usenames,dvipsnames,table]{xcolor}
\usepackage{booktabs}

\newcommand{\probP}{\text{I\kern-0.15em P}}
\begin{document}

\title{\large\bfseries Transfer entropy for finite data}

\author{Alec Kirkley}
\email{alec.w.kirkley@gmail.com}
\affiliation{Institute of Data Science, University of Hong Kong, Hong Kong SAR, China}
\affiliation{Department of Urban Planning and Design, University of Hong Kong, Hong Kong SAR, China}
\affiliation{Urban Systems Institute, University of Hong Kong, Hong Kong SAR, China}

\date{\today}

\begin{abstract}
Transfer entropy is a widely used measure for quantifying directed information flows in complex systems. While the challenges of estimating transfer entropy for continuous data are well known, it has two major shortcomings for data of finite cardinality: it exhibits a substantial positive bias for sparse bin counts, and it has no clear means to assess statistical significance. By computing information content in finite data streams without explicitly considering symbols as instances of random variables, we derive a transfer entropy measure which is asymptotically equivalent to the standard plug-in estimator but remedies these issues for time series of small size and/or high cardinality, permitting a fully nonparametric assessment of statistical significance without simulation.  
\end{abstract}

\maketitle

Transfer entropy \cite{schreiber2000measuring} provides a general technique for assessing the extent to which one temporal dynamics is capable of forecasting another \cite{hamilton2020time}, serving as a flexible nonlinear alternative to the celebrated Granger causality \cite{granger1969investigating,barnett2009granger}. Consequently, transfer entropy has seen a broad variety of applications, from inferring effective connectivity among regions of the brain \cite{vicente2011transfer} to identifying chains of influence in armed conflicts \cite{kushwaha2023discovering}. Despite its popularity, it is widely acknowledged that the estimation of transfer entropy in empirical data is challenging due to its high dimensionality, sensitivity to noise, and underlying causality assumptions \cite{smirnov2013spurious,james2016information,bossomaier2016transfer}. A number of improved transfer entropy estimators for continuous data have thus been proposed, which utilize robust differential entropy estimation \cite{paninski2003estimation,kraskov2004estimating}, relative signal ordering \cite{staniek2008symbolic}, Lempel-Ziv complexity \cite{restrepo2020transfer}, and graphical models \cite{runge2012escaping}.  

But transfer entropy has two fundamental flaws for time series of \emph{finite} length and cardinality, a case receiving less attention. First, it rarely indicates a null effect (transfer entropy of zero) in practice due to the usage of Shannon conditional entropy, which is always reduced by conditioning on additional variables \cite{cover1991elements}. This non-negativity is often preferred in information theoretic measures, but it necessarily leads to a positivity bias in which data generated from uncorrelated random processes have a non-negligible amount of shared information \cite{zhang2015evaluating}. The second fundamental flaw is that statistical significance of transfer entropy values must be assessed by permutation testing or bootstrapping through simulations \cite{marschinski2002analysing,dimpfl2013using}. These approaches are not only computationally intensive but require the choice of a significance level, which needs correction for multiple comparisons when constructing networks from time series \cite{novelli2019large} and arguably should be adjusted based on the length of the time series \cite{lindley1957statistical}.

Here we derive a combinatorial transfer entropy measure---which we call the \emph{reduced} transfer entropy---that alleviates both of the above issues at no additional computational cost by avoiding the standard interpretation of the observed time series as realizations of random variables. Our measure also allows for automatic selection of the optimal temporal lag, and can be generalized to compute multivariate transfer entropy.

\textit{Transfer entropy}---Consider a pair of scalar-valued time series $\bm{x},\bm{y}\in \{1,...,C\}^{T}$ of finite length $T$ in discretized time that each have the same cardinality of $C$ possible values, which we will generically call ``symbols'' \cite{cover1991elements}. For continuous time series, a cardinality of $C$ may be achieved through some discretization or density estimation process if direct entropy estimation is not employed \cite{paninski2003estimation,kraskov2004estimating}. Let $k,l<T-1$ be temporal lags for the series $\bm{x}$ and $\bm{y}$ respectively, and 
\begin{align}
\bm{x}^{(-k)}_t &= [x_t,x_{t-1},...,x_{t-k+1}]   \\
\bm{y}^{(-l)}_t &= [y_t,y_{t-1},...,y_{t-l+1}]  
\end{align}
be the $k$- and $l$-dimensional delay embedding vectors for $\bm{x}$ and $\bm{y}$ at time $t$, respectively \cite{packard1980geometry}. Also let $\bm{X}^{(-k)}=\{\bm{x}^{(-k)}_t\}_{t=T-N}^{T-1}$ and $\bm{Y}^{(-l)}=\{\bm{y}^{(-l)}_t\}_{t=T-N}^{T-1}$ be the $N\times k$- and $N\times l$-dimensional matrices containing these embeddings for all timesteps, with $N=T-\text{max}(k,l)$ the number of timesteps available for computing the embeddings. Finally, denote with $\bm{y}^{(+1)}=\{y_{t+1}\}_{t=T-N}^{T-1}$ the shifted time series $\bm{y}$ one time step forward. The standard ``plug-in'' transfer entropy estimator from $\bm{x}$ to $\bm{y}$ under the lag specification $(k,l)$ is then given by
\begin{align}\label{eq:TE-mostgeneral}
\mathcal{T}^{(k,l)}_{\bm{x}\to\bm{y}} = 
H_S(\bm{y}^{(+1)}\vert \bm{Y}^{(-l)}) - H_S(\bm{y}^{(+1)}\vert \bm{Z}^{(-k,-l)}), 
\end{align}
where $\bm{Z}^{(-k,-l)} = [\bm{Y}^{(-l)}\mid\mid\bm{X}^{(-k)}]$ is the $N\times (l+k)$-dimensional delay embedding matrix concatenating $\bm{Y}^{(-l)}$ and $\bm{X}^{(-k)}$, and
\begin{align}\label{eq:ShannonCEprobs}
H_S(\bm{w}\vert \bm{V})&= \!-\!\sum_{w_t,\bm{v}_t}P(w_t,\bm{v}_t)\log \frac{P(w_t,\bm{v}_t)}{P(\bm{v}_t)} \\
&= -\frac{1}{N}\sum_{r,\bm{s}}n^{(\bm{w},\bm{V})}_{r,\bm{s}}\log \frac{n^{(\bm{w},\bm{V})}_{r,\bm{s}}}{n^{(\bm{V})}_{\bm{s}}}   
\label{eq:ShannonCEcounts}
\end{align}
is the Shannon conditional entropy of a scalar time series $\bm{w}=\{w_t\}$ given a vector time series $\bm{V}=\{\bm{v}_t\}$ with the same temporal indices. Here, 
\begin{align}\label{eq:contingency}
n^{(\bm{w},\bm{V})}_{r,\bm{s}} = \sum_{t}\delta(w_t,r)\delta(\bm{v}_t,\bm{s}) 
\end{align}
is the number of timesteps at which the two series---which may in general be scalar- or vector-valued---take the particular value combination $(r,\bm{s})$, and
\begin{align}\label{eq:margin}
n^{(\bm{V})}_{\bm{s}} = \sum_{r}n^{(\bm{w},\bm{V})}_{r,\bm{s}} = \sum_{t}\delta(\bm{v}_t,\bm{s}) 
\end{align}
is the number of occurrences of $\bm{s}$ in $\bm{V}$. One can store all of these joint counts in the \emph{contingency table} $\bm{n}^{(\bm{w},\bm{V})}=\{n^{(\bm{w},\bm{V})}_{r,\bm{s}}\}_{r,\bm{s}}$, which fully specifies the empirical joint probability distribution over symbol combinations. (We also use the notation $\log_2\equiv \log$ for brevity.) 

Eq.~\ref{eq:TE-mostgeneral} computes the average number of additional bits we can save in specifying a future value $y_{t+1}$ of  $\bm{y}$ when we encode it using the past values $\bm{x}^{(-k)}_t$ of $\bm{x}$ and $\bm{y}^{(-l)}_t$ of $\bm{y}$. The parameters $k,l$ are chosen using domain expertise or, in some cases, general time series model selection techniques \cite{barnett2012transfer}. By using the Shannon conditional entropy of Eq.~\ref{eq:ShannonCEprobs}, Eq.~\ref{eq:TE-mostgeneral} quantifies information sharing among the underlying \emph{probability distributions} assumed to generate the observed time series data, which are estimated using plug-in estimators.

As discussed, Eq.~\ref{eq:TE-mostgeneral} seldom indicates conditional independence of $\bm{x}$ and $\bm{y}$, i.e. a transfer entropy of exactly zero, and we require computationally expensive simulations with a potentially arbitrary choice of significance level to determine whether a measured transfer entropy value is meaningful \cite{marschinski2002analysing,dimpfl2013using}. These issues are fundamentally a result of treating $\bm{x},\bm{y}$ as random variables, since the standard Shannon entropy $H_S(\bm{y}^{(+1)}\vert \bm{Y}^{(-l)})$ is always greater than $H_S(\bm{y}^{(+1)}\vert \bm{Z}^{(-k,-l)})$. By considering a \emph{combinatorial} formulation in which shared information among the embeddings $\{\bm{y}^{(+1)},\bm{Y}^{(-l)},\bm{X}^{(-k)}\}$ is calculated
for a single transmission event, we can avoid explicitly imposing any distributional assumptions. This enables us to circumvent the positivity bias and automatically determine statistical significance using the Minimum Description Length (MDL) principle \cite{rissanen1978modeling}.   

\emph{Combinatorial conditional entropy}---Consider a sender, Alice, who wants to transmit a scalar-valued time series $\bm{w}\in \{1,..,C\}^N$ to a receiver, Bob, with the help of a different (vector-valued) time series $\bm{V}\in \{1,..,C\}^{N\times d}$ which is already known by both parties. In order to exploit Bob's knowledge of $\bm{V}$ to transmit $\bm{w}$, Alice must tell him the shared structure among $\bm{V}$ and $\bm{w}$, since this shared structure will constrain the possibilities for $\bm{w}$ and consequently reduce the number of bits needed for its transmission. The overlap among $\bm{w}$ and $\bm{V}$ can be summarized with a contingency table $\bm{n}^{(\bm{w},\bm{V})}$ as in Eq.~\ref{eq:contingency}.

Since Bob knows $\bm{V}$, Alice only needs to encode contingency tables $\bm{n}^{(\bm{w},\bm{V})}$ that satisfy the margin constraint of Eq.~\ref{eq:margin} for all $\bm{s}$, the sum of the row indexed by $\bm{s}$ in the contingency table. There are ${n^{(\bm{V})}_{\bm{s}}+C-1\choose C-1}$ ways to assign $C$ non-negative integer values to the $\bm{s}$-th row that sum to $n^{(\bm{V})}_{\bm{s}}$. Thus, taking a product over the row indices $\bm{s}$, there are
\begin{align}\label{eq:Omega}
\Omega(\bm{n}^{(\bm{w},\bm{V})}\vert \bm{n}^{(\bm{V})})
 = \prod_{\bm{s}}{n^{(\bm{V})}_{\bm{s}}+C-1\choose C-1}
\end{align}
total possibilities for the contingency table given the margin constraints Bob already knows. Alice can then construct a fixed length code over these contingency tables with codelength $\log\Omega$ \cite{cover1991elements}. This forms the first contribution to the conditional entropy. 

Now that Bob knows the series overlap stored in the contingency table $\bm{n}^{(\bm{w},\bm{V})}$, Alice can send him $\bm{w}$ at a lower information cost. Since he knows the $n_{\bm{s}}^{(\bm{V})}$ locations at which $\bm{V}$ takes the value $\bm{s}$, Alice can construct a fixed-length code with a codelength of 
\begin{align}\label{eq:CEMicro}
H_M(\bm{w}\vert \bm{V}) = \frac{1}{N}\log \prod_{\bm{s}}\frac{n_{\bm{s}}^{(\bm{V})}!}{\prod\limits_{r}n_{r\bm{s}}^{(\bm{w},\bm{V})}!}
\end{align}
bits per timestep. (In other words, codelength divided by $N$, for consistency with the Shannon formulation in Eq.~\ref{eq:ShannonCEcounts}.) This is just the logarithm of the number of timestep assignments for $r=1,...,C$ in $\bm{w}$ consistent with the contingency table. As lag embeddings are not statistically independent due to their necessary overlap, there are symbol permutations that do not correspond to possible time series. Therefore, the proposed encoding has redundancy according to the Kraft-McMillan inequality \cite{mackay2003information}, providing an upper bound on the combinatorial entropy. Nevertheless, Eq.~\ref{eq:CEMicro} naturally equates to the Shannon conditional entropy in Eq.~\ref{eq:ShannonCEcounts} when we apply the Stirling approximation $\log n! \approx n\log n - n/\ln(2)$.

Putting both contributions together, the coding rate under this transmission scheme is
\begin{align}\label{eq:CE-corrected}
H_C(\bm{w}\vert \bm{V}) 
&=\frac{\log\Omega(\bm{n}^{(\bm{w},\bm{V})}\vert \bm{n}^{(\bm{V})})}{N}+ H_M(\bm{w}\vert \bm{V}) 
\end{align}
bits per timestep for Alice to transmit $\bm{w}$ to Bob given their shared knowledge of $\bm{V}$. Eq.~\ref{eq:CE-corrected} is thus a measure of conditional entropy between $\bm{w}$ and $\bm{V}$ that highlights the finite nature of the time series by no longer considering them as random variables but as complete populations to be transmitted only once. Applying Stirling's approximation to Eq.~\ref{eq:CE-corrected}, we obtain
\begin{align}\label{eq:stirling-similar}
H_C(\bm{w}\vert \bm{V}) \approx \frac{\log\Omega}{N} + H_S(\bm{w}\vert \bm{V}),      
\end{align}
where $H_S(\bm{w}\vert \bm{V})$ is the standard Shannon conditional entropy of Eq.~\ref{eq:ShannonCEcounts}. For $N\to\infty$ and fixed $C$, the correction $\log \Omega/N$ vanishes asymptotically (see Supplemental Material \cite{SM}) and $H_C\approx H_S$. However, in practice the difference between the two expressions results in substantial discrepancies in the transfer entropy.

\emph{Reduced transfer entropy}---Using Eq.~\ref{eq:CE-corrected} we can define a transfer entropy for finite time series $\bm{x}$ and $\bm{y}$:
\begin{align}\label{eq:HMminusHM}
\!\!\!\!\mathcal{R}&^{(k,l)}_{\bm{x}\to \bm{y}} = H_C(\bm{y}^{(+1)}\vert \bm{Y}^{({-}l)})\!-\!H_C(\bm{y}^{(+1)}\vert \bm{Z}^{({-}k,{-}l)})\\
&=\Delta^{(k,l)}_{\bm{x}\to \bm{y}} + \frac{1}{N}\log \frac{\prod\limits_{q,\bm{r},\bm{s}}n^{(+1,{-}l,{-}k)}_{q,\bm{r},\bm{s}}!\prod\limits_{\bm{r}}n^{({-}l)}_{\bm{r}}!}{\prod\limits_{q,\bm{r}}n^{({+}1,-l)}_{q,\bm{r}}!\prod\limits_{\bm{r},\bm{s}}n^{({-}l,{-}k)}_{\bm{r},\bm{s}}!}, \label{eq:reduced-TE}
\end{align}
where
\begin{align}\label{eq:correction}
\Delta^{(k,l)}_{\bm{x}\to \bm{y}} = \frac{1}{N}\log \frac{\prod\limits_{\bm{r}}\bigbinom{n^{(-l)}_{\bm{r}}+C-1}{C-1}}{\prod\limits_{\bm{r},\bm{s}}\bigbinom{n^{({-}l,{-}k)}_{\bm{r},\bm{s}}+C-1}{C-1}},    
\end{align}
and
\begin{align}\label{eq:n3pt}
n&^{(+1,{-}l,{-}k)}_{q,\bm{r},\bm{s}} = n^{(\bm{y}^{(+1)},\bm{Y}^{(-l)},\bm{X}^{(-k)})}_{q,\bm{r},\bm{s}} \nonumber\\
&= \sum_{t=T-N}^{T-1}\delta(y_{t+1},q)\delta(\bm{y}^{({-}l)}_t,\bm{r})\delta(\bm{x}^{({-}k)}_t,\bm{s})   \end{align}
is a multi-dimensional contingency table, with the other terms in Eq.~\ref{eq:reduced-TE} giving its marginal sums. We call Eq.~\ref{eq:reduced-TE} the ``reduced'' transfer entropy since one can prove that the correction satisfies $\Delta \leq 0$ (see Supplemental Material).

\begin{figure}[t]
    \centering
    \includegraphics[width=\columnwidth]{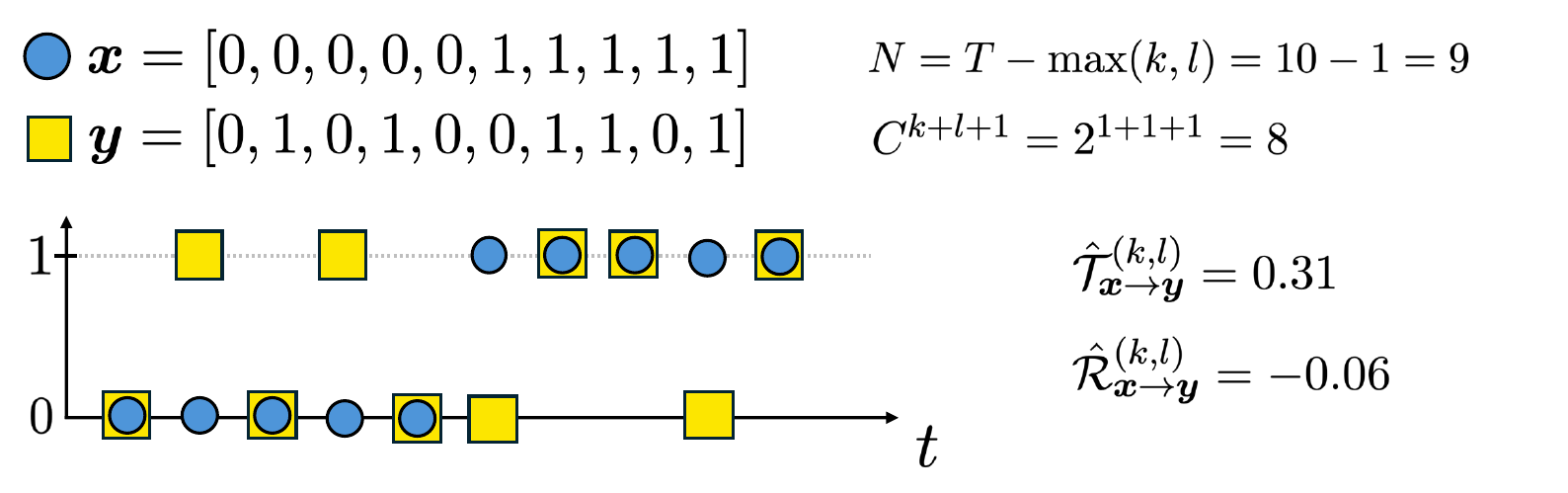}
    \caption{Transfer entropy of random time series with sparse counts and lags $k=l=1$. Sparsity here is from small $T$, but can also arise from large lags $k,l$ or cardinality $C$. A negative reduced transfer entropy indicates that it is more compressive to transmit $\bm{y}$'s future values $\bm{y}^{(+1)}$ using only its own past values $\bm{Y}^{(-l)}$ than to also include $\bm{x}$'s past values $\bm{X}^{(-k)}$.}
    \label{fig:main-diagram}
\end{figure}

To naturally address the issue of statistical significance, the reduced transfer entropy of Eq.~\ref{eq:reduced-TE} allows for negative values. A negative reduced transfer entropy happens precisely when the additional cost of including $\bm{X}^{(-k)}$ is greater than the amount of information we save when using it to transmit $\bm{y}^{(+1)}$. We are thus effectively performing model selection using the Minimum Description Length (MDL) principle \cite{rissanen1978modeling}, which is equivalent to Bayesian model selection using joint model probabilities under appropriate choices of likelihood and priors \cite{mackay2003information}. As we show in our experiments, this provides an alternative to frequentist permutation testing with no need for expensive simulations or the choice of a significance level. The reduced transfer entropy can be calculated with negligible additional computational cost since Eq.~\ref{eq:Omega} can be computed directly from the contingency table. Moreover, this formulation allows for selection of the optimal lags $k,l$ according to the MDL principle: the MDL-optimal lags $k,l$ are those that produce the highest value of the reduced transfer entropy (see Supplemental Material). Additionally, as the combinatorial conditional entropy in Eq.~\ref{eq:CEMicro} accounts for the information to specify the joint dependencies among two time series, it more heavily penalizes using sparsely observed symbol combinations for compression. This helps mitigate the positivity bias of Eq.~\ref{eq:TE-mostgeneral} as $T$ decreases, the lags $k,l$ increase, or the cardinality $C$ increases (see Supplemental Material).

\begin{figure}[t]
    \centering
    \includegraphics[width=\columnwidth]{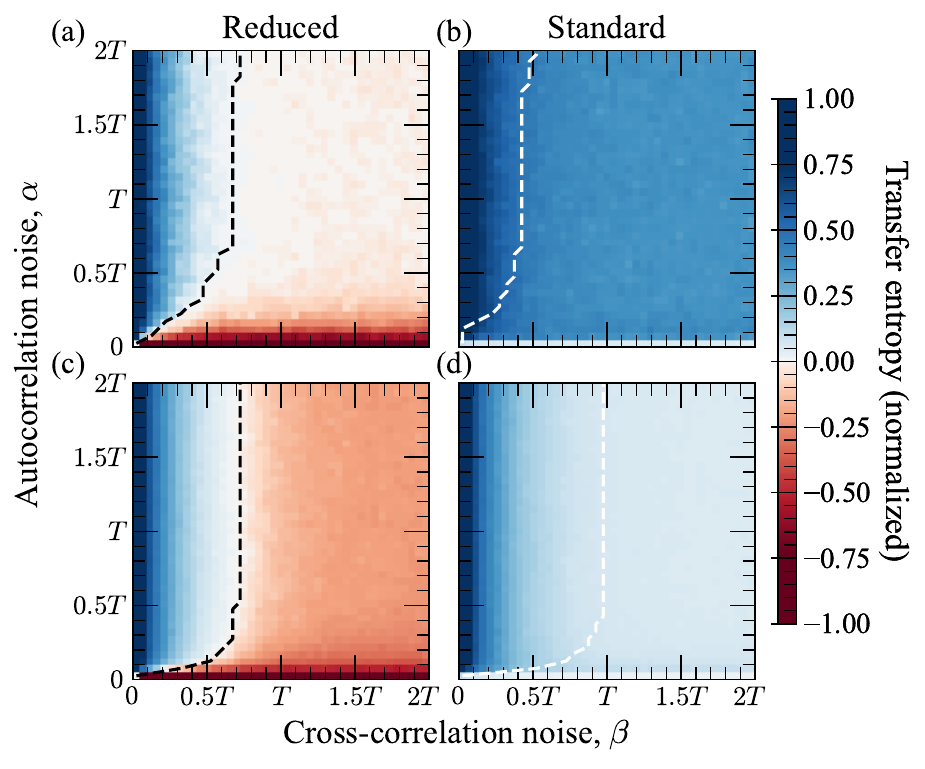}
    \caption{Transfer entropy of synthetic time series. (a) Normalized reduced transfer entropy (Eq.~\ref{eq:normTEreduced}) versus cross- and auto-correlation noise, for synthetic time series $\bm{x},\bm{y}$ with $\{T,l,C\}=\{100,3,2\}$. The region of statistical significance ($\mathcal{R}>0$) is indicated with a black line. (b) Normalized standard transfer entropy (Eq.~\ref{eq:normTEstandard}) for the same time series, statistical significance ($p<0.05$ from permutation testing) indicated with a white line. Panels (c) and (d) repeat these experiments for $T=1000$.}
    \label{fig:main-heat}
\end{figure}

It is worth noting that there are alternative finite-size corrections proposed for mutual information, which could in principle also be used to compute transfer entropy. Multiple works have suggested subtracting off an expectation value of mutual information over a suitable null model \cite{vinh2009information,zhang2015evaluating,lai2016corrected}, although this approach also lacks a clear model selection mechanism---an observed effect can be greater than one expects by chance, but it is not necessarily \emph{significantly} greater. Even in asymptotic regimes where higher moments \cite{hutter2001distribution}---and thus a $z$-score---can be computed for certain null models, these methods require the choice of a significance level. 

A few existing methods avoid these issues by considering multi-step encodings in the computation of mutual information \cite{newman2020improved,jerdee2024mutual}. The goal of these measures is to compute an \emph{unconditional} mutual information, which results in encodings that are not well suited for constructing a transfer entropy measure. The method of \cite{newman2020improved} imposes a symmetry in the encoding that allows for a symmetric mutual information measure but assumes a potentially substantial amount of additional information is known for the target vector ahead of time (the marginal symbol counts) than assumed by our encoding. This measure additionally requires an approximation of the entropy of the contingency table \cite{jerdee2024improved}. Meanwhile, the encoding in \cite{jerdee2024mutual} requires optimization over a free parameter, resulting in an inconsistent encoding among the two conditional entropy expressions in Eq.~\ref{eq:HMminusHM}, as well as unclear upper and lower bounds. In contrast, the encoding proposed above is specifically designed to exploit the inherent asymmetry of transfer entropy for an efficient encoding that is consistent across the different conditioning variables to enable clear bounds and normalization.

\emph{Normalization}---To have an absolute scale on which to interpret transfer entropies, we can normalize both the standard transfer entropy $\mathcal{T}$ (Eq.~\ref{eq:TE-mostgeneral}) and the reduced transfer entropy $\mathcal{R}$ (Eq.~\ref{eq:reduced-TE}) by the maximum value they can attain over all possible $\bm{x}$. Using the bounds obtained in the Supplemental Material, we can construct normalized measures
\begin{align}\label{eq:normTEstandard}
\hat{\mathcal{T}}^{(k,l)}_{\bm{x}\to \bm{y}} &= \frac{\mathcal{T}^{(k,l)}_{\bm{x}\to \bm{y}}}{H_S(\bm{y}^{(+1)}\vert \bm{Y}^{(-l)})},\\ 
\hat{\mathcal{R}}^{(k,l)}_{\bm{x}\to \bm{y}} &= \frac{\mathcal{R}^{(k,l)}_{\bm{x}\to \bm{y}}}{M^{(k,l)}_{\bm{x},\bm{y}}}\label{eq:normTEreduced},
\end{align}
where $M^{(k,l)}_{\bm{x},\bm{y}}={-}\Delta^{(k,l)}_{\bm{x}\to \bm{y}}$ for $\mathcal{R}^{(k,l)}_{\bm{x}\to \bm{y}} \leq 0$ and $M^{(k,l)}_{\bm{x},\bm{y}}=\Delta^{(k,l)}_{\bm{x}\to \bm{y}}+H_M(\bm{y}^{(+1)}\vert \bm{Y}^{(-l)})$ for $\mathcal{R}^{(k,l)}_{\bm{x}\to \bm{y}} > 0$.

This choice of normalization maps to $\hat{\mathcal{T}}^{(k,l)}_{\bm{x}\to \bm{y}}\in [0,1]$ and $\hat{\mathcal{R}}^{(k,l)}_{\bm{x}\to \bm{y}}\in [-1,1]$, with both lower bounds saturated if and only if we are completely certain about the value of $\bm{x}_t^{(-k)}=\bm{s}$ when $\bm{y}_t^{(-l)}=\bm{r}$ is known, or equivalently when $\bm{x}_t^{(-k)}$ and $\bm{y}_t^{(-l)}$ provide redundant information and are identical up to symbol relabelling (i.e. bin permutation). Both upper bounds are saturated when we are completely certain about the value $y_{t+1}=q$ when $\bm{x}^{(-k)}_t=\bm{s}$ and $\bm{y}^{(-l)}_t=\bm{r}$ are both known, meaning that $\bm{x}^{(-k)}_t$ maximally reduces the uncertainty about $y_{t}$ given the information provided by $\bm{y}^{(-l)}_t$. (See Supplemental Material for details.) This allows us to analyze the transfer entropy measures on comparable absolute scales---with $\hat{\mathcal{T}}^{(k,l)}_{\bm{x}\to \bm{y}}=0$ corresponding to $\hat{\mathcal{R}}^{(k,l)}_{\bm{x}\to \bm{y}}=-1$ and $\hat{\mathcal{T}}^{(k,l)}_{\bm{x}\to \bm{y}}=1$ corresponding to $\hat{\mathcal{R}}^{(k,l)}_{\bm{x}\to \bm{y}}=1$---while allowing $\hat{\mathcal{R}}^{(k,l)}_{\bm{x}\to \bm{y}}>0$ to signify significance according to the MDL principle.

\begin{figure}[b]
    \centering
    \includegraphics[width=\linewidth]{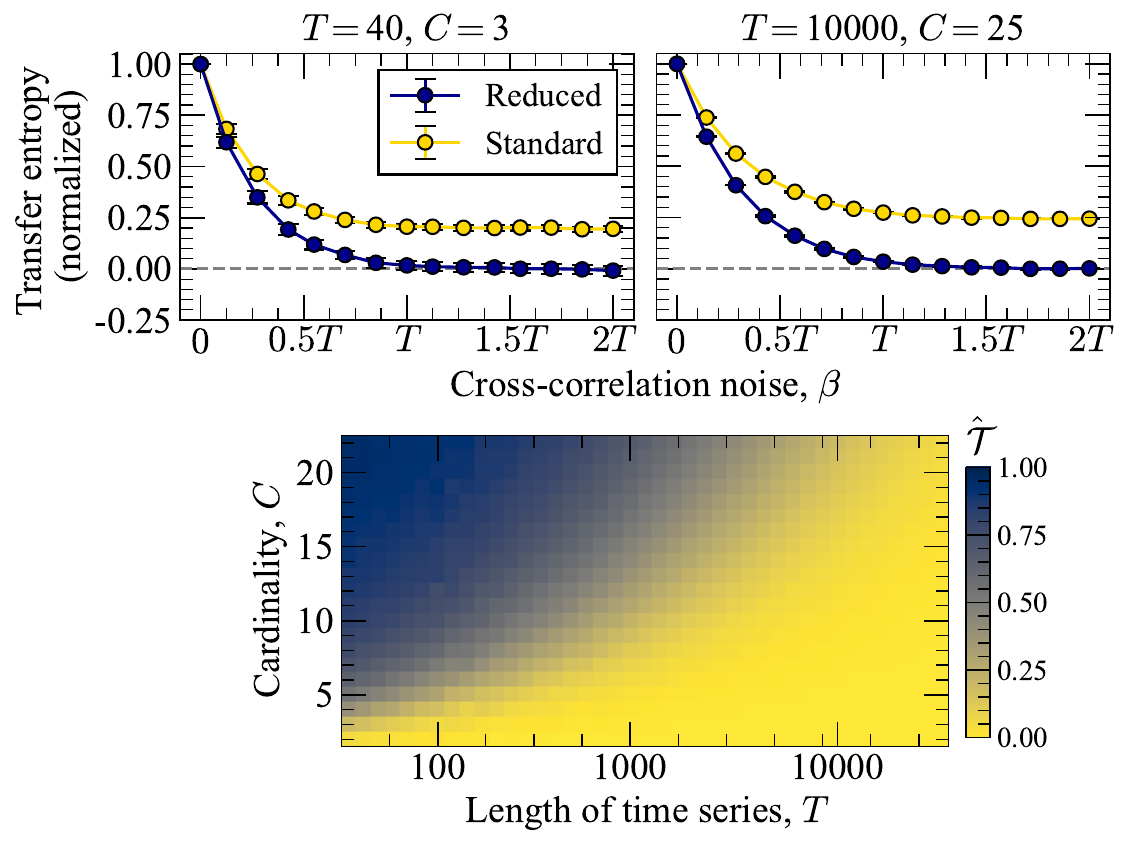}
    \caption{Positivity bias of transfer entropy. (Top) Normalized transfer entropies versus cross-correlation noise, $\beta$, for synthetic time series with $(T,C)=(40,3),(10000,25)$ and $k=l=1$. (Bottom) Standard transfer entropy as a function of $T,C$ for time series $\bm{x},\bm{y}\in \{1,...,C\}^{T}$ generated uniformly at random, with $k=l=1$.}
    \label{fig:cross-section-main}
\end{figure}

\textit{Tests on synthetic data}---
To test our method in a controlled setting, we generate synthetic time series pairs $\bm{x},\bm{y}$ with tunable auto- and cross-correlation. We first generate the time series $\bm{y}$ as $\lfloor T/l\rfloor$ copies of a vector of length $l$ drawn uniformly at random from $\{1,...,C\}$---this initializes $\bm{y}$ to have perfect autocorrelation at lag $l$, so that $\bm{Y}^{(-l)}$ provides complete information about $\bm{y}^{(+1)}$. We then shuffle $\bm{y}$ a number of times $\alpha$ to add noise to the autocorrelation. After this, we generate $\bm{x}$ as a copy of the shuffled $\bm{y}$ shifted back by $l$ timesteps (with periodic boundaries), so that $\bm{X}^{(-l)}$ provides complete information about $\bm{y}^{(+1)}$. Finally, we shuffle $\bm{x}$ a number of times $\beta$ to add noise to the cross-correlation among $\bm{x}$ and $\bm{y}$. Fig.~\ref{fig:main-heat} shows how the reduced (left) and standard (right) transfer entropy measures behave as the two sources of noise $\alpha$ and $\beta$ are varied for $T\in \{100,1000\}$ (top and bottom rows, respectively), with a lag of $l=3$ and cardinality $C=2$. Results are averaged over $200$ simulations. 

Both the reduced and standard measures find statistical significance at high $\alpha$ and low $\beta$---in this regime, $\bm{Y}^{(-l)}$ provides very noisy information about $\bm{y}^{(+1)}$ while $\bm{X}^{(-l)}$ is highly cross-correlated with $\bm{y}^{(+1)}$. However, the reduced measure is able to detect this statistical significance without any simulations or significance level. Additionally, while the region of significance remains consistent for the reduced measure as $T$ is increased, it becomes larger for the standard measure, which has to rely on a permutation test with a pre-specified significance level (here, $p=0.05$) to detect significance. This is a common issue for frequentist methods, which easily report significance for larger sample sizes but lack a principled significance level adjustment \cite{lindley1957statistical}. We can also see a substantial positive bias in the standard transfer entropy for $T=100$ (panel (b)), which takes a normalized value of $\approx 0.35$ in the high noise regime. This bias becomes less apparent for $T=1000$, but is still noticeable, with a normalized value of $\approx 0.05$. 

In Fig.~\ref{fig:cross-section-main} we repeat a similar set of experiments but remove autocorrelation by generating $\bm{y}\in \{1,...,C\}^{T}$ uniformly at random. We can see that both transfer entropy measures steadily decrease as noise is added, but that the standard transfer entropy levels off at a value much higher than zero ($\approx 0.20-0.25$) in the high noise regime for both small $T$ (top left) and large $C$ (top right). The reduced measure corrects for this positive bias as expected. Error bars here represent two standard errors in the mean over $200$ simulations at each value of $\beta$. In the bottom panel we show the bias in the standard normalized transfer entropy as a function of $T,C$ for $k=l=1$ by computing its average value over $200$ simulated pairs of completely uncorrelated random time series $\bm{x},\bm{y}\in \{1,...,C\}^{T}$. We find confirmation that the bias worsens as $T$ decreases and as $C$ increases. The bias is also inflated at larger lags $k,l$ (see Supplementary Material).

\begin{figure}[t]
    \centering
    \includegraphics[width=1\linewidth]{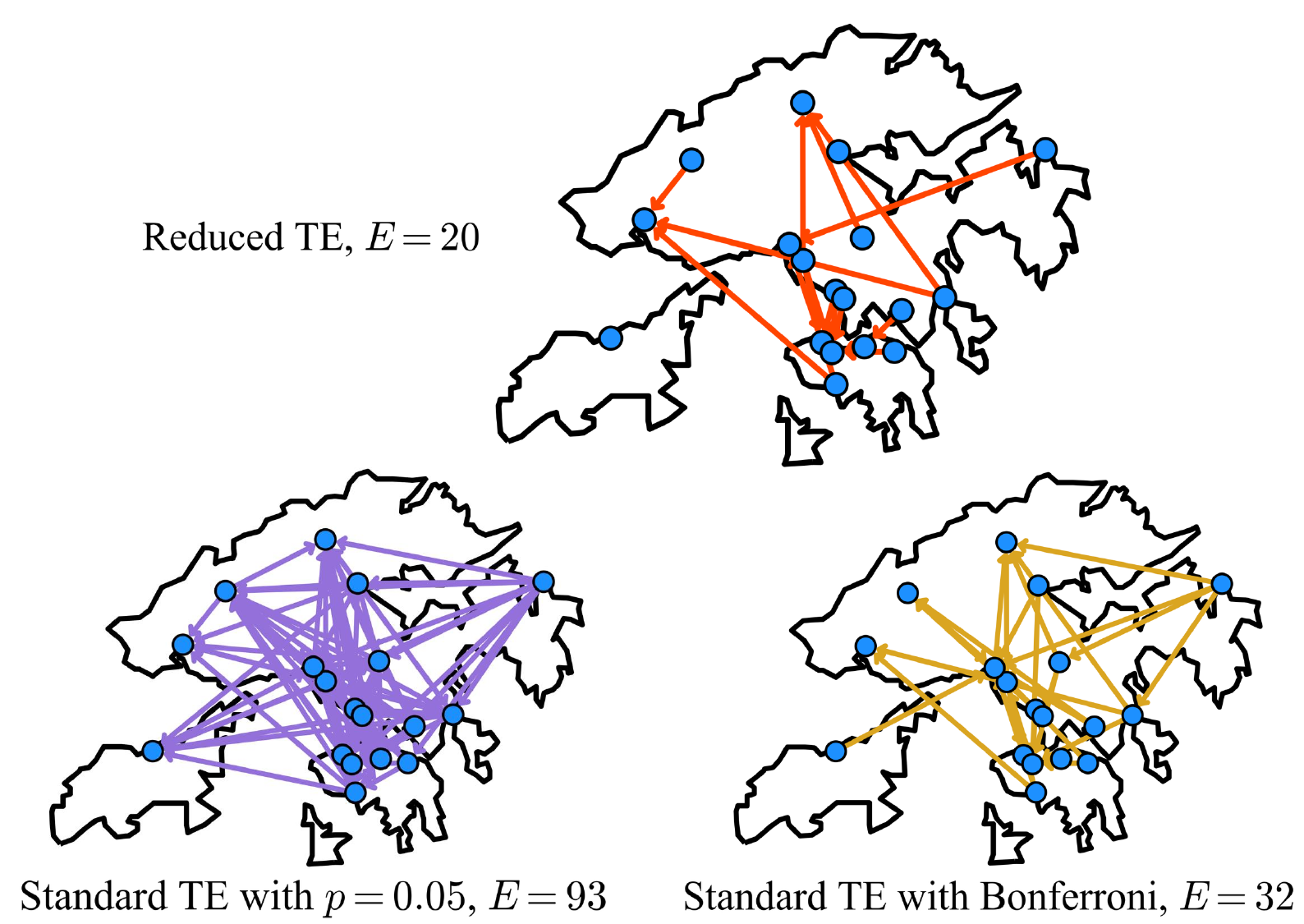}
    \caption{The transfer entropy correction can substantially impact downstream results. Networks were constructed from hourly sampled time series recording air quality across Hong Kong at 18 different measurement stations (nodes). We place a directed edge $(i,j)$ if the time series $\bm{x}$ at $i$ has a statistically significant transfer entropy with the series $\bm{y}$ at $j$ with $k=l=1\text{ hour}$. The number of edges $E$ in the final network are listed above each panel.}
    \label{fig:HK}
\end{figure}

\textit{Air pollution case study}---As it captures general nonlinear dependencies in spatiotemporal data, transfer entropy has been employed widely in the climate and atmospheric sciences \cite{runge2012escaping,hu2022multi,tongal2022transfer}. Here we demonstrate the impact the transfer entropy reduction can have for real time series data by examining time series of the Air Quality Health Index (AQHI) \cite{wong2013developing} across Hong Kong, which categorizes overall health risk due to different air pollutants on a categorical scale $\{\text{low, moderate, high, very high, serious}\}$, forming the bins for our analyses. Data were collected for April 2025, which had substantial fluctuations in air quality from a set of dust storms. Directed transfer entropy networks were constructed using the reduced transfer entropy---a positive value indicating statistical significance---as well as the standard transfer entropy with a significance threshold set to $p=0.05$, with and without Bonferroni correction for multiple comparisons, for permutation tests with $1000$ trials. The results are shown in Fig.~\ref{fig:HK}, where we can see major differences in the inferred significant links across the three methods. Thus, using the reduced transfer entropy, one could arrive at qualitatively different conclusions about the causal dependencies needed for models of urban air pollution \cite{luo2025air} or other environmental dynamics  \cite{gelbrecht2017complex,runge2019inferring}.

\emph{Conclusion}---Here we derive a combinatorial transfer entropy measure for finite data which corrects for the positivity bias of standard transfer entropy and allows for automatic model selection using the MDL principle. It will be important for future research to examine how this measure changes conclusions in various applications inferring information flows in complex systems, and to extend the measure using different encodings that can compress asynchronous time series and other structural signatures in real-world data. Code implementing the reduced transfer entropy can be found at \url{https://github.com/aleckirkley/Reduced-Transfer-Entropy}.


%
%

\clearpage

\onecolumngrid


\begin{center}
  \textbf{\large Supplemental Material for: \\ \vspace{0.25cm}
Transfer entropy of finite time series \vspace{0.25cm}} \\[.2cm]

  Alec Kirkley,$^{1,2,3}$ \\ [.1cm]
  {\itshape ${}^1$Institute of Data Science, University of Hong Kong, Hong Kong SAR, China \\
            ${}^2$Department of Urban Planning and Design, University of Hong Kong, Hong Kong SAR, China \\
            ${}^3$Urban Systems Institute, University of Hong Kong, Hong Kong SAR, China \\
            }
\end{center}

\setcounter{equation}{0}
\setcounter{figure}{0}
\setcounter{table}{0}
\setcounter{page}{1}
\setcounter{section}{0}
\renewcommand{\theequation}{S\arabic{equation}}
\renewcommand{\thefigure}{S\arabic{figure}}
\renewcommand{\thetable}{S\arabic{table}}
\renewcommand{\thepage}{S\arabic{page}}
\renewcommand{\thesection}{S\arabic{section}}
\renewcommand{\bibnumfmt}[1]{[#1]}
\renewcommand{\citenumfont}[1]{#1}

\title{\large\bfseries Supplementary Material for: Transfer entropy for finite data}

\author{Alec Kirkley}
\email{alec.w.kirkley@gmail.com}
\affiliation{Institute of Data Science, University of Hong Kong, Hong Kong SAR, China}
\affiliation{Department of Urban Planning and Design, University of Hong Kong, Hong Kong SAR, China}
\affiliation{Urban Systems Institute, University of Hong Kong, Hong Kong SAR, China}

\date{\today}

\maketitle

\section{Asymptotic equivalence of $H_S$ and $H_M$}

Using Eq.s~4~and~9, we have
\begin{align}
N\times H_S(\bm{w}\vert \bm{V}) &= \sum_{\bm{s}}n_{\bm{s}}\log n_{\bm{s}} - \sum_{r,\bm{s}}n_{r,\bm{s}}\log n_{r,\bm{s}}, \\
N\times H_M(\bm{w}\vert \bm{V}) &= \sum_{\bm{s}}\log n_{\bm{s}}! - \sum_{r,\bm{s}}\log n_{r,\bm{s}}!,
\end{align}
where we have omitted superscripts for brevity. In the limit $n_{r,\bm{s}}\gg 1$ for all $r,\bm{s}$, we can use the Stirling approximations
\begin{align}
\log (n_{r,\bm{s}}!) &\approx n_{r,\bm{s}}\log n_{r,\bm{s}}-n_{r,\bm{s}}/\ln(2),\\
\log (n_{\bm{s}}!) &\approx n_{\bm{s}}\log n_{\bm{s}}-n_{\bm{s}}/\ln(2),
\end{align}
giving
\begin{align}
N\times H_M(\bm{w}\vert \bm{V}) &\approx 
\sum_{\bm{s}}[n_{\bm{s}}\log n_{\bm{s}}-n_{\bm{s}}/\ln(2)] - \sum_{r,\bm{s}}[n_{r,\bm{s}}\log n_{r,\bm{s}}-n_{r,\bm{s}}/\ln(2)]\\
&=\sum_{\bm{s}}n_{\bm{s}}\log n_{\bm{s}} - N/\ln(2) - \sum_{r,\bm{s}}n_{r,\bm{s}}\log n_{r,\bm{s}} + N/\ln(2)\\
&= N\times H_S(\bm{w}\vert \bm{V}),
\end{align}
and thus $H_S\approx H_M$ in this regime.

\section{Asymptotic scaling of conditional entropy correction $\log\Omega/N$}

From Eq.~8, we have
\begin{align}
\frac{\log \Omega}{N} &= \frac{1}{N}\sum_{\bm{s}}\log{n^{(\bm{V})}_{\bm{s}}+C-1\choose C-1}\\
&\leq \frac{1}{N}\sum_{\bm{s}}\log \frac{(n^{(\bm{V})}_{\bm{s}}+C-1)^{(C-1)}}{(C-1)!}\\
&\leq \frac{1}{N}\sum_{\bm{s}}(C-1)\log (N+C-1)\\
&\leq \frac{\log (N+C-1)}{N}(C-1)C^k,
\end{align}
where $k$ is the dimension (i.e. lag in the transfer entropy calculation) of the time series $\bm{V}$. In the limit $N\to\infty$ with $C$ fixed, this expression vanishes and $H_C\approx H_M \approx H_S$.

\section{Non-positivity of transfer entropy correction $\Delta$}

The correction to the transfer entropy in Eq.~14 is given by
\begin{align}\label{eq:corrmultiset}
\Delta^{(k,l)}_{\bm{x}\to \bm{y}} = \frac{1}{N}\log \frac{\prod\limits_{\bm{r}}\bigbinom{n^{(\bm{Y}^{(-l)})}_{\bm{r}}+C-1}{C-1}}{\prod\limits_{\bm{r},\bm{s}}\bigbinom{n^{(\bm{Y}^{(-l)},\bm{X}^{(-k)})}_{\bm{r},\bm{s}}+C-1}{C-1}}
= \frac{1}{N}\log \prod\limits_{\bm{r}}\frac{\multiset{C}{n^{(\bm{Y}^{(-l)})}_{\bm{r}}}}{\prod\limits_{\bm{s}}\multiset{C}{n^{(\bm{Y}^{(-l)},\bm{X}^{(-k)})}_{\bm{r},\bm{s}}}},
\end{align}
where
\begin{align}
\multiset{n}{k} = {n+k-1\choose k}    
\end{align}
is the multiset coefficient counting the number of unique ways to construct a multiset of $k$ elements with draws from set of $n$ elements with replacement. In \cite{felippe2024network} it was shown that multiset coefficients satisfy 
\begin{align}
\multiset{n}{k}\multiset{n}{l} \geq \multiset{n}{k+l},  
\end{align}
for any $n,k,l$. This inequality implies that
\begin{align}
\multiset{C}{n^{(\bm{Y}^{(-l)})}_{\bm{r}}} \leq \prod\limits_{\bm{s}}\multiset{C}{n^{(\bm{Y}^{(-l)},\bm{X}^{(-k)})}_{\bm{r},\bm{s}}},    
\end{align}
since 
\begin{align}
\sum_{\bm{s}}n^{(\bm{Y}^{(-l)},\bm{X}^{(-k)})}_{\bm{r},\bm{s}} = n^{(\bm{Y}^{(-l)})}_{\bm{r}}.    
\end{align}
Thus, we have
\begin{align}
\Delta^{(k,l)}_{\bm{x}\to \bm{y}}
 = \frac{1}{N}\sum_{\bm{r}}\log \frac{\multiset{C}{n^{(\bm{Y}^{(-l)})}_{\bm{r}}}}{\prod\limits_{\bm{s}}\multiset{C}{n^{(\bm{Y}^{(-l)},\bm{X}^{(-k)})}_{\bm{r},\bm{s}}}}
 \leq \frac{1}{N}\sum_{\bm{r}}\log \frac{\prod\limits_{\bm{s}}\multiset{C}{n^{(\bm{Y}^{(-l)},\bm{X}^{(-k)})}_{\bm{r},\bm{s}}}}{\prod\limits_{\bm{s}}\multiset{C}{n^{(\bm{Y}^{(-l)},\bm{X}^{(-k)})}_{\bm{r},\bm{s}}}}
 =\frac{1}{N}\sum_{\bm{r}}\log(1) = 0,
\end{align}
and the correction is bounded above by zero. Hence the proposed transfer entropy measure is a ``reduced'' version of the standard transfer entropy.

\section{Positive bias of the standard transfer entropy}
Since it is strictly non-negative, any statistical fluctuations must push the standard transfer entropy of Eq.~4 towards values greater than zero for uncorrelated time series, for which we would ideally return zero to indicate a lack of forecasting capability of $\bm{y}$ from $\bm{x}$. We refer to this as the ``positive bias'' of the transfer entropy, and here try to understand how it behaves analytically.

For completely uncorrelated time series, the true underlying probability of each outcome $(q,\bm{r},\bm{s})$ is $1/C^{k+l+1}$. Thus, under a maximum entropy assumption the contingency table $\bm{n}$ will be distributed like a multinomial distribution with uniform bin probabilities $1/C^{k+l+1}$. The expected value of the transfer entropy under this null model is  
\begin{align}
\expec{\mathcal{T}^{(k,l)}_{\bm{x}\to\bm{y}}} &= 
\expec{H_S(\bm{y}^{(+1)}\vert \bm{Y}^{(-l)})} - \expec{H_S(\bm{y}^{(+1)}\vert \bm{Z}^{(-k,-l)})}\\
&=\expec{H_{q,\bm{r}}} + \expec{H_{\bm{r},\bm{s}}} -\expec{H_{\bm{r}}} - \expec{H_{q,\bm{r},\bm{s}}}, 
\end{align}
where
\begin{align}
\expec{H_{\bm{a}}} &= -\left\langle\sum_{\bm{a}}\frac{n_{\bm{a}}}{N}\log \frac{n_{\bm{a}}}{N}\right\rangle \\
&=\frac{1}{N}\left\langle\sum_{\bm{a}}n_{\bm{a}}\log N - n_{\bm{a}}\log n_{\bm{a}}\right\rangle \\
&=\log N - \frac{1}{N}\left\langle\sum_{\bm{a}}n_{\bm{a}}\log n_{\bm{a}}\right\rangle \\
&=\log N - \frac{\abs{\mathcal{X}(\bm{a})}}{N}\left\langle n_{\bm{a}}\log n_{\bm{a}}\right\rangle
\end{align}
is the expected Shannon entropy of counts indexed by $\bm{a}$ when the multi-way contingency table $\bm{n}$ of all joint counts is distributed as a uniform multinomial, and $\abs{\mathcal{X}(\bm{a})}$ is the size of $\bm{a}$'s support. This holds since when the full multi-way contingency table entries $\{n_{q,\bm{r},\bm{s}}\}$ are uniform-multinomially distributed, all of the table marginals are also uniform-multinomially distributed over the corresponding new smaller dimension. Now, the marginal distribution of $n_{\bm{a}}$ is a Binomial with $N$ trials and success probability $1/\abs{\mathcal{X}(\bm{a})}$, so the desired expectation is given by
\begin{align}
\left\langle n_{\bm{a}}\log n_{\bm{a}}\right\rangle
 = \sum_{k=0}^{N}{N\choose k}\left(\frac{1}{\abs{\mathcal{X}(\bm{a})}}\right)^{k}\left(1-\frac{1}{\abs{\mathcal{X}(\bm{a})}}\right)^{N-k}k\log k,
\end{align}
which unfortunately is intractable. However, if we approximate the binomial distribution as a Normal distribution in the limit of large $N$ and fixed $\abs{\mathcal{X}(\bm{a})}$, we have that the expectation is approximately
\begin{align}
\left\langle n_{\bm{a}}\log n_{\bm{a}}\right\rangle
 &\approx \mathbb{E}_{x\sim \mathcal{N}(\mu,\sigma^2)}[x\log x],
\end{align}
where 
\begin{align}
\mu &= \frac{N}{\abs{\mathcal{X}(\bm{a})}},\\
\sigma^2 &= \frac{N}{\abs{\mathcal{X}(\bm{a})}}\left(1-\frac{1}{\abs{\mathcal{X}(\bm{a})}}\right).
\end{align}
This is also intractable but admits a simple solution if we expand $x\log x$ via Taylor expansion about the maximum, thus 
\begin{align}
\mathbb{E}[x\log x] &\approx \mathbb{E}\left[\mu \log \mu + (1+\log \mu)(x-\mu) + \frac{1}{2\mu}(x-\mu)^2 \right] \\
&= \mu \log \mu + \frac{\sigma^2}{2\mu}\\
&= \frac{N}{\abs{\mathcal{X}(\bm{a})}} \log \frac{N}{\abs{\mathcal{X}(\bm{a})}} + \frac{1}{2}\left(1-\frac{1}{\abs{\mathcal{X}(\bm{a})}}\right).
\end{align}
We thus have
\begin{align}
\expec{H_{\bm{a}}} = \log \abs{\mathcal{X}(\bm{a})} - \frac{\abs{\mathcal{X}(\bm{a})}}{2N}\left(1-\frac{1}{\abs{\mathcal{X}(\bm{a})}}\right).    
\end{align}
Subbing this in to our expression for the expected transfer entropy and replacing the $\abs{\mathcal{X}(\bm{a})}$ terms appropriately, we have
\begin{align}
\expec{\mathcal{T}^{(k,l)}_{\bm{x}\to\bm{y}}} 
&=\expec{H_{q,\bm{r}}} + \expec{H_{\bm{r},\bm{s}}} -\expec{H_{\bm{r}}} - \expec{H_{q,\bm{r},\bm{s}}}\\
&\approx \frac{C^l}{2N}\left(1-\frac{1}{C^l}\right)
+ \frac{C^{k+l+1}}{2N}\left(1-\frac{1}{C^{k+l+1}}\right)
- \frac{C^{l+1}}{2N}\left(1-\frac{1}{C^{l+1}}\right)
- \frac{C^{k+l}}{2N}\left(1-\frac{1}{C^{k+l}}\right)\\
&= \frac{C^l(C^k-1)(C-1)}{2N}\\
&\sim \frac{C^{k+l+1}}{N}.
\end{align}

Looking at this result, we can see that the positive bias of the standard transfer entropy will become greater as: (1) the cardinality $C$ increases; (2) the number of timesteps $T$ decreases; or (3) the lags $k,l$ increase. This is consistent with the results seen in Figures~1~and~2 in the main text. All of these factors lead to sparser bin counts, which intuitively should result in less reliable estimation. However, in the same regime of uniformity, the reduced transfer entropy gives a downward adjustment of 
\begin{align}
\Delta^{(k,l)}_{\bm{x}\to \bm{y}}
 = \frac{1}{N}\sum_{\bm{r}}\log \frac{\multiset{C}{n^{(\bm{Y}^{(-l)})}_{\bm{r}}}}{\prod\limits_{\bm{s}}\multiset{C}{n^{(\bm{Y}^{(-l)},\bm{X}^{(-k)})}_{\bm{r},\bm{s}}}}
 \sim \frac{C^l}{N}\log \frac{\multiset{C}{N/C^l}}{\multiset{C}{N/C^{k+l}}^{C^{k}}}
 \sim O\left(\frac{C^{k+l+1}\log N}{N}\right).
\end{align}
Thus, the reduced transfer entropy helps to correct the positive bias of the standard transfer entropy. This is also consistent with what is observed in our numerical experiments.

\section{Bounds on transfer entropies}

From Eq.~4, we find
\begin{align}
\mathcal{T}^{(k,l)}_{\bm{x}\to\bm{y}} &= 
H_S(\bm{y}^{(+1)}\vert \bm{Y}^{(-l)}) - H_S(\bm{y}^{(+1)}\vert \bm{Z}^{(-k,-l)}) \leq H_S(\bm{y}^{(+1)}\vert \bm{Y}^{(-l)}),
\end{align}
since the conditional entropies are both non-negative. Thus, $H_S(\bm{X}^{(-k)}\vert \bm{Y}^{(-l)})$ is an upper bound on $\mathcal{T}^{(k,l)}_{\bm{x}\to\bm{y}}$ over all time series $\bm{x}$. Similarly, for the reduced transfer entropy we have
\begin{align}
\mathcal{R}^{(k,l)}_{\bm{x}\to\bm{y}} &= 
\Delta^{(k,l)}_{\bm{x}\to \bm{y}} + \frac{1}{N}\log \frac{\prod\limits_{q,\bm{r},\bm{s}}n^{(+1,{-}l,{-}k)}_{q,\bm{r},\bm{s}}!\prod\limits_{\bm{r}}n^{({-}l)}_{\bm{r}}!}{\prod\limits_{q,\bm{r}}n^{({+}1,-l)}_{q,\bm{r}}!\prod\limits_{\bm{r},\bm{s}}n^{({-}l,{-}k)}_{\bm{r},\bm{s}}!}\\
&\leq \Delta^{(k,l)}_{\bm{x}\to \bm{y}} + \frac{1}{N}\log \frac{\prod\limits_{\bm{r},\bm{s}}n^{({-}l,{-}k)}_{\bm{r},\bm{s}}!\prod\limits_{\bm{r}}n^{({-}l)}_{\bm{r}}!}{\prod\limits_{q,\bm{r}}n^{({+}1,-l)}_{q,\bm{r}}!\prod\limits_{\bm{r},\bm{s}}n^{({-}l,{-}k)}_{\bm{r},\bm{s}}!}\\
&= \Delta^{(k,l)}_{\bm{x}\to \bm{y}} + \frac{1}{N}\log \frac{\prod\limits_{\bm{r}}n^{({-}l)}_{\bm{r}}!}{\prod\limits_{q,\bm{r}}n^{({+}1,-l)}_{q,\bm{r}}!}\\
&= \Delta^{(k,l)}_{\bm{x}\to \bm{y}} + H_M(\bm{y}^{(+1)}\vert \bm{Y}^{(-l)}),
\end{align}
giving $\Delta^{(k,l)}_{\bm{x}\to \bm{y}} + H_M(\bm{y}^{(+1)}\vert \bm{Y}^{(-l)})$ as an upper bound on $\mathcal{R}^{(k,l)}_{\bm{x}\to\bm{y}}$. Both of these upper bounds are saturated when
\begin{align}\label{eq:uppersaturated}
n^{({-}l,{-}k)}_{\bm{r},\bm{s}}  
 = n^{(+1,{-}l,{-}k)}_{q,\bm{r},\bm{s}} 
\end{align}
for all $\bm{r},\bm{s}$ and some $q$ fixed by $\bm{r},\bm{s}$. In other words, this upper bound is saturated when we are completely certain about the value $y_{t+1}=q$ when $\bm{x}^{(-k)}_t=\bm{s}$ and $\bm{y}^{(-l)}_t=\bm{r}$ are both known, meaning that $\bm{x}^{(-k)}_t$ maximally reduces the uncertainty about $y_{t+1}$ given the information provided by $\bm{y}^{(-l)}_t$.  

For lower bounds, the standard transfer entropy satisfies $\mathcal{T}^{(k,l)}_{\bm{x}\to\bm{y}}\geq 0$, as conditioning always reduces entropy. Meanwhile, for the reduced transfer entropy we have
\begin{align}
\mathcal{R}^{(k,l)}_{\bm{x}\to\bm{y}} &= 
\Delta^{(k,l)}_{\bm{x}\to \bm{y}} + H_M(\bm{y}^{(+1)}\vert \bm{Y}^{(-l)})-H_M(\bm{y}^{(+1)}\vert \bm{Y}^{(-l)},\bm{X}^{(-k)}). 
\end{align}
The term $H_M(\bm{y}^{(+1)}\vert \bm{Y}^{(-l)})$ counts ($1/N$ times the logarithm of) the number of valid configurations of $\bm{y}^{(+1)}$ given the constraints provided by $\bm{Y}^{(-l)}$ and the contingency table $\bm{n}^{(\bm{y}^{(+1)},\bm{Y}^{(-l)})}$. Meanwhile, the term $H_M(\bm{y}^{(+1)}\vert \bm{Y}^{(-l)},\bm{X}^{(-k)})$ counts ($1/N$ times the logarithm of) the number of valid configurations of $\bm{y}^{(+1)}$ given the constraints provided by $\bm{Y}^{(-l)}$, $\bm{X}^{(-k)}$, and the contingency table $\bm{n}^{(\bm{y}^{(+1)},\bm{Y}^{(-l)},\bm{X}^{(-k)})}$. There must be at least as many valid configurations of $\bm{y}^{(+1)}$ under the first set of constraints as the second, since any valid configuration of $\bm{y}^{(+1)}$ under the second set of constraints is also valid under the first set of constraints. This is the combinatorial equivalent to conditioning always reducing entropy. Therefore, we have
\begin{align}
H_M(\bm{y}^{(+1)}\vert \bm{Y}^{(-l)}) \geq  H_M(\bm{y}^{(+1)}\vert \bm{Y}^{(-l)},\bm{X}^{(-k)})    
\end{align}
and so 
\begin{align}
\mathcal{R}^{(k,l)}_{\bm{x}\to\bm{y}} = 
\Delta^{(k,l)}_{\bm{x}\to \bm{y}} + H_M(\bm{y}^{(+1)}\vert \bm{Y}^{(-l)}) -  H_M(\bm{y}^{(+1)}\vert \bm{Y}^{(-l)},\bm{X}^{(-k)})
\geq \Delta^{(k,l)}_{\bm{x}\to \bm{y}}.
\end{align}
Thus, $\Delta^{(k,l)}_{\bm{x}\to \bm{y}}$ is a lower bound for $\mathcal{R}^{(k,l)}_{\bm{x}\to\bm{y}}$. Both of these lower bounds are saturated when 
\begin{align}\label{eq:lowersaturated}
n^{({-}l)}_{\bm{r}} = n^{({-}l,{-}k)}_{\bm{r},\bm{s}}
\end{align}
for all $\bm{r}$ and some $\bm{s}$ fixed by $\bm{r}$. In other words, this lower bound is saturated when we are completely certain about the value of $\bm{x}_t^{(-k)}=\bm{s}$ when $\bm{y}_t^{(-l)}=\bm{r}$ is known, or equivalently when $\bm{x}_t^{(-k)}$ and $\bm{y}_t^{(-l)}$ provide redundant information as they are identical up to relabelings of symbols.

Although it is a valid upper bound, $\Delta^{(k,l)}_{\bm{x}\to \bm{y}} + H_M(\bm{X}^{(-k)}\vert \bm{Y}^{(-l)})$ may be negative or zero for $\mathcal{R}^{(k,l)}_{\bm{x}\to\bm{y}}\leq 0$, in which case it does not provide a suitable normalization for $\hat{\mathcal{R}}^{(k,l)}_{\bm{x}\to\bm{y}}$ as it does not allow us to uniquely determine the sign of $\mathcal{R}$ from its normalized value. Therefore, when $\mathcal{R}^{(k,l)}_{\bm{x}\to\bm{y}}\leq 0$, we can use the alternative upper bound $-\Delta^{(k,l)}_{\bm{x}\to \bm{y}}\geq 0$, so that the minimum reduced transfer entropy is found at $\mathcal{R}^{(k,l)}_{\bm{x}\to\bm{y}}=\Delta^{(k,l)}_{\bm{x}\to \bm{y}}$ with a normalized value of $\hat{\mathcal{R}}^{(k,l)}_{\bm{x}\to\bm{y}}=-1$ in Eq.~17. $-\Delta^{(k,l)}_{\bm{x}\to \bm{y}}$ provides a valid normalization for $\mathcal{R}\leq 0$ since in this case we have
\begin{align}
\abs{\mathcal{R}^{(k,l)}_{\bm{x}\to\bm{y}}} = -\mathcal{R}^{(k,l)}_{\bm{x}\to\bm{y}}=  -\Delta^{(k,l)}_{\bm{x}\to \bm{y}} -[ H_M(\bm{y}^{(+1)}\vert \bm{Y}^{(-l)}) -  H_M(\bm{y}^{(+1)}\vert \bm{Y}^{(-l)},\bm{X}^{(-k)})]
\leq -\Delta^{(k,l)}_{\bm{x}\to \bm{y}}.
\end{align}
The final step follows from $H_M(\bm{y}^{(+1)}\vert \bm{Y}^{(-l)}) \geq  H_M(\bm{y}^{(+1)}\vert \bm{Y}^{(-l)},\bm{X}^{(-k)})$.

\section{Finite-size deviation between $H_S$ and $H_M$}

As shown, the Shannon conditional entropy $H_S$ (Eq.~5) and the microcanonical conditional entropy $H_M$ (Eq.~9) are equivalent when $n_{r,\bm{s}}\gg 1$ for all $r,\bm{s}$. However, as discussed, we are often not in this count-rich regime in practice. Therefore, $H_S$ and $H_M$ may differ considerably. Outside of the count-rich regime, we must take higher order terms in the Stirling approximation
\begin{align}
\log n! \approx n\log n-n/\ln(2)+\frac{1}{2}\log(2\pi n),    
\end{align}
giving
\begin{align}
N\times (H_M-H_S) &= \sum_{\bm{s}}[\log n_{\bm{s}}! - n_{\bm{s}}\log n_{\bm{s}}]
+ \sum_{r,\bm{s}}[n_{r,\bm{s}}\log n_{r,\bm{s}}-\log n_{r,\bm{s}}!]\\
&\approx \frac{1}{2}\sum_{\bm{s}}\log (2\pi n_{\bm{s}})
- \frac{1}{2}\sum_{r,\bm{s}}\log (2\pi n_{r,\bm{s}}).
\end{align}
This suggests that the discrepancy between $H_S$ and $H_M$ vanishes when $n_{\bm{s}}=n_{r,\bm{s}}$ for all $\bm{s}$ and some single $r$ for each $\bm{s}$. But this is simply when $H_S=H_M=0$ are both minimized. 

On the other hand, if we are in the regime with
\begin{align}
n_{r,\bm{s}}=\frac{n_{\bm{s}}}{\abs{\mathcal{X}(r)}},    
\end{align}
this will be the regime in which the conditional entropies are maximized, since we learn nothing about $r$ by knowing $\bm{s}$ when $n_{r,\bm{s}}=n_{r',\bm{s}}$ for all $r,r',\bm{s}$. In this case, we have
\begin{align}
N\times (H_M-H_S) &\approx \frac{1}{2}\sum_{\bm{s}}\log (2\pi n_{\bm{s}})
- \frac{1}{2}\sum_{r,\bm{s}}\log (2\pi n_{\bm{s}}/\abs{\mathcal{X}(r)}).
\end{align}
This can be simplified further when the counts $n_{\bm{s}}=N/\abs{\mathcal{X}(\bm{s})}$ are equal, giving
\begin{align}
N\times (H_M-H_S) &\approx \frac{1}{2}\sum_{\bm{s}}\log (2\pi N/\abs{\mathcal{X}(\bm{s})})
- \frac{1}{2}\sum_{r,\bm{s}}\log (2\pi N/\abs{\mathcal{X}(\bm{s})}/\abs{\mathcal{X}(r)})\\
&=\frac{\abs{\mathcal{X}(\bm{s})}}{2}\log \frac{2\pi N}{\abs{\mathcal{X}(\bm{s})}} - \frac{\abs{\mathcal{X}(r)}\abs{\mathcal{X}(\bm{s})}}{2}\log \frac{2\pi N}{\abs{\mathcal{X}(r)}\abs{\mathcal{X}(\bm{s})}}\\
&=\frac{\abs{\mathcal{X}(\bm{s})}}{2}\log \frac{2\pi N}{\abs{\mathcal{X}(\bm{s})}}[1-\abs{\mathcal{X}(r)}]
+ \frac{\abs{\mathcal{X}(r)}\abs{\mathcal{X}(\bm{s})}}{2}\log \abs{\mathcal{X}(r)}.
\end{align}
This expression suggests that, as $N$ grows with $\abs{\mathcal{X}(r)},\abs{\mathcal{X}(\bm{s})}$ fixed, we have multiple distinct regimes:
\begin{enumerate}
    \item For $N \lesssim \abs{\mathcal{X}(r)}\abs{\mathcal{X}(\bm{s})}$, $H_M$ exceeds $H_S$, with a considerable discrepancy of at least $\frac{\abs{\mathcal{X}(r)}\abs{\mathcal{X}(\bm{s})}}{2}\log \abs{\mathcal{X}(r)}$ when $N\lesssim \abs{\mathcal{X}(\bm{s})}$
    \item For $N \gtrsim \abs{\mathcal{X}(r)}\abs{\mathcal{X}(\bm{s})}$, we have that $H_S$ exceeds $H_M$. 
\end{enumerate}
Thus, the difference between $H_S$ and $H_M$ is not consistent for finite data and may impact the transfer entropy differently depending on count sparsity. This is a further reason to use  the reduced measure, as it is adapted specifically to the finite-size case as it does not utilize Stirling's approximation.

\section{Multivariate reduced transfer entropy}
One can extend the proposed transfer entropy measure to the multivariate case, in which we condition the entropy of the future time series values $\bm{y}^{(+1)}$ on multiple additional time series embeddings $\bm{W}^{(-l)}=[\bm{W}_1^{(-l)}\vert\vert\bm{W}_2^{(-l)}\vert\vert \cdots]$, to determine whether a candidate time series $\bm{X}^{(-l)}$ provides meaningful information about the target $\bm{y}^{(+1)}$ beyond $\bm{W}^{(-l)}$ and $\bm{Y}^{(-l)}$. Multivariate transfer entropy is a popular method for constructing networks from time series \cite{runge2012escaping,novelli2019large}. Given a larger set of time series that form the nodes of a network, we can determine which subset of time series $\partial_i=\{\bm{x}_{j_1},\bm{x}_{j_2},\cdots\}$ provide complementary information for predicting future values of a target time series $\bm{x}_i$, with time series in $\partial_i$ becoming the in-neighbors of the node corresponding to $\bm{x}_i$. The multivariate transfer entropy formulation ensures that each additional neighbor added to $\partial_i$ provides meaningful new information about $\bm{x}_i$, beyond what the existing neighbors provide, to avoid redundancies.

Keeping consistency with the standard definition, the multivariate reduced transfer entropy $\mathcal{M}^{(l)}_{\bm{x}\to \bm{y}\vert \bm{W}}$ for time series $\bm{x}$ and $\bm{y}$ conditioned on other time series embeddings $\bm{W}$ is given by
\begin{align}\label{eq:MVTE}
\mathcal{M}^{(l)}_{\bm{x}\to \bm{y}\vert \bm{W}} = H_C(\bm{y}^{(+1)}\vert [\bm{Y}^{(-l)}\vert\vert\bm{W}^{(-l)}]) - H_C(\bm{y}^{(+1)}\vert [\bm{Z}^{(-l,-l)}\vert\vert \bm{W}^{(-l)}]),    
\end{align}
where $\bm{Z}^{(-l,-l)}$ is defined as before and $\bm{W}^{(-l)}=[\bm{W}_1^{(-l)}\vert\vert\bm{W}_2^{(-l)}\vert\vert \cdots]$ is the concatenation of an arbitrary number of embeddings from other time series. One can easily utilize different lags $k,l,\cdots$ here as well for each time series, but for notational simplicity we use a single lag $l$. We can compute Eq.~\ref{eq:MVTE} easily by recognizing that
\begin{align}
\mathcal{M}^{(l)}_{\bm{x}\to \bm{y}\vert \bm{W}} = \mathcal{R}^{(l,l)}_{[\bm{x}\vert\vert \bm{W}]\to \bm{y}} - \mathcal{R}^{(l,l)}_{\bm{W}\to \bm{y}},      
\end{align}
where $[\bm{x}\vert\vert \bm{W}]$ is the joint time series formed by $\bm{x}$ and the time series used to form $\bm{W}$, and $\mathcal{R}$ is the original (bivariate) reduced transfer entropy.

\clearpage
\section{Additional tests for synthetic time series}

In this supplement we include additional experimental results applying the transfer entropy measures to synthetic time series.

\begin{figure}[htb!]
    \centering
    \includegraphics[width=0.8\linewidth]{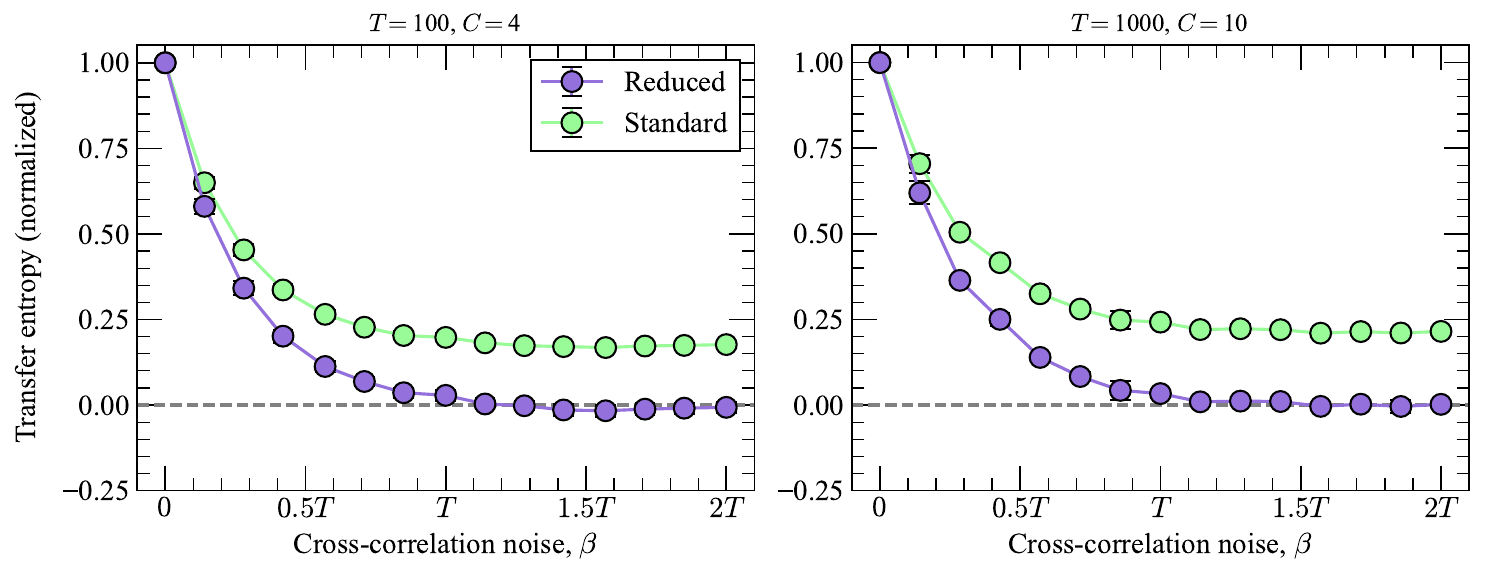}
    \caption{(Top) Standard and reduced (normalized) transfer entropy versus cross-correlation noise, $\beta$, for synthetically generated time series $\bm{x},\bm{y}$ with $(T,C)=(100,4)$. (Bottom) Same plot for $(T,C)=(1000,10)$. A lag of $k=1$ is imposed between $\bm{x}$ and $\bm{y}$.}
    \label{fig:cross-sec-appendix}
\end{figure}

\begin{figure}[htb!]
    \centering
    \includegraphics[width=0.7\linewidth]{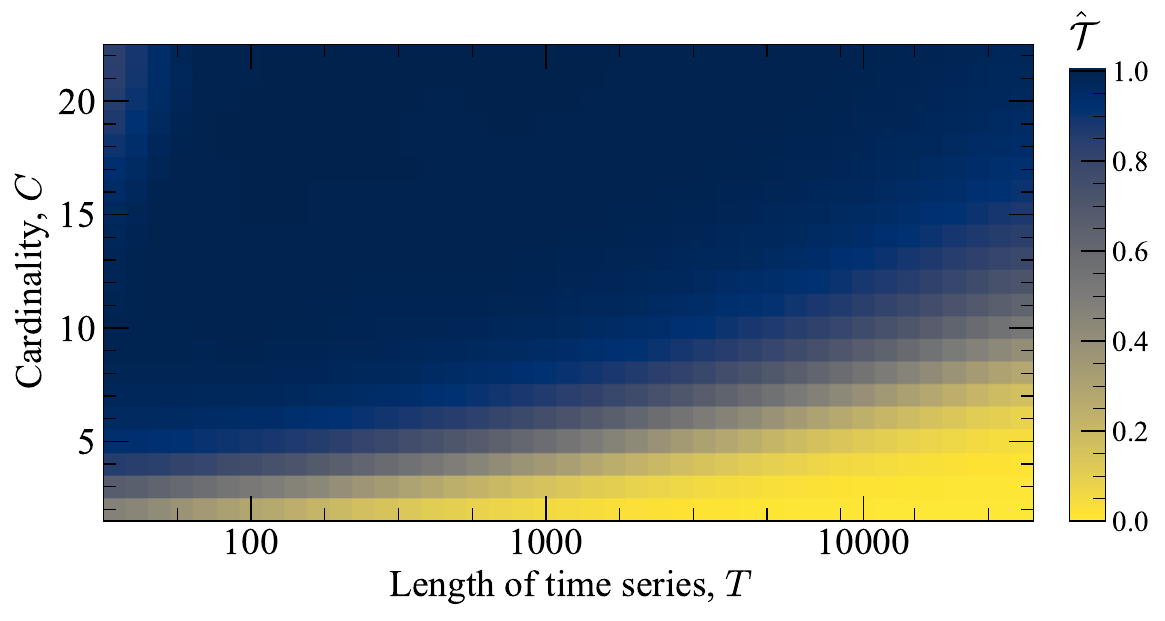}
    \caption{Standard normalized transfer entropy (Eq.~16) versus cardinality $C$ and length $T$ for completely uncorrelated random pairs of time series, with lags set to $k=l=2$. We can see an amplification of the effect shown in Fig.~3 in the main text due to the further sparsification of the bin counts from the larger embedding dimension.}
    \label{fig:TCscan-appendix}
\end{figure}

In Figures~\ref{fig:cross-sec-appendix}~and~\ref{fig:TCscan-appendix}, we repeat the analyses of Figure~3 in the main text, for different choices of cardinality $C$ and number of timesteps $T$. We find consistent results with those of the main text, with an amplification of the bias in the standard transfer entropy due to the larger choice of embedding dimension $l$, since the number of bins in the contingency table $\bm{n}^{(+1,-l,-l)}$ increases with $l$.

\begin{figure}[htb!]
    \centering
    \includegraphics[width=1\linewidth]{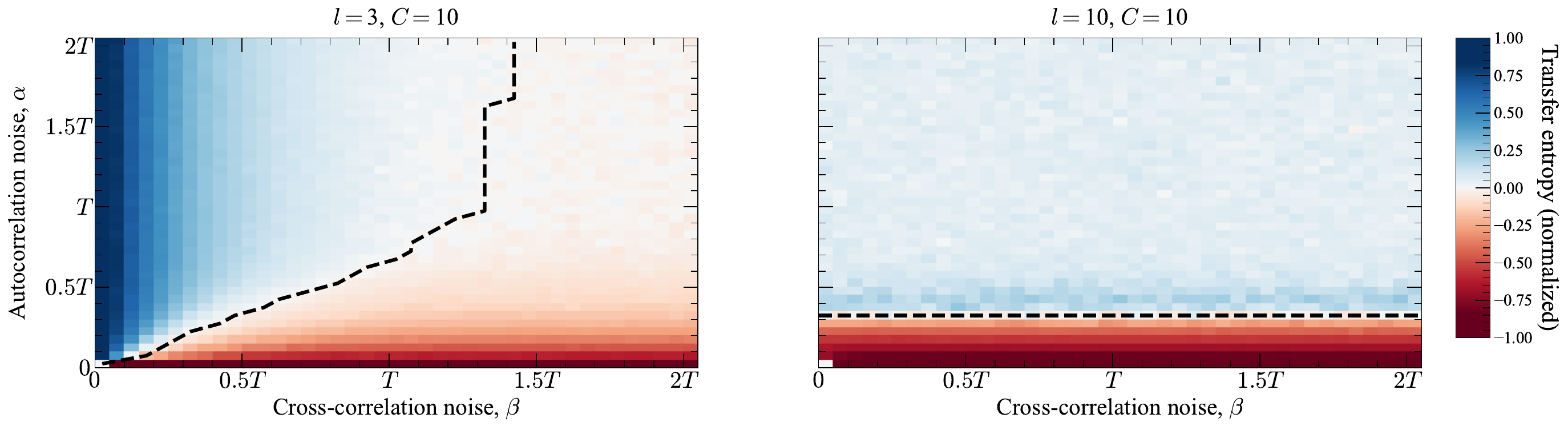}
    \caption{(Left) Normalized reduced transfer entropy  versus cross- and auto-correlation noise, for synthetic time series with $\{T, l, C\}$ = $\{1000, 3, 10\}$. The region of statistical significance (positive reduced transfer entropy) is indicated with a black line. (Right) Same experiment, but now with $l=10$.}
    \label{fig:heatmap-appendix}
\end{figure}

In Fig.~\ref{fig:heatmap-appendix}, we repeat the analyses of Figure 2 in the main text, for higher embedding dimension $l$ and higher cardinality $C$, which further sparsifies the bin counts and amplifies the positivity bias that needs to be corrected for by the reduced transfer entropy. In the left panel, we increase the cardinality $C$ from $C=2$ to $C=10$, where we can see generally the same pattern as in Figure 2. We observe slightly more cross-correlation noise required for a null result due to a weaker autocorrelation in $\bm{y}$ at any given value of $\alpha$, from the greater cardinality of the symbol set. In the right panel, we also increase the lag dimension to $l=10$, which has a more substantial effect on the results due to the number of unique bins in the contingency $\bm{n}^{(+1,-l,-l)}$ table scaling as $C^{l+l+1}$. Here we see that, due to the long lag and high cardinality, the autocorrelation of $\bm{y}$ is weakened to the point that  any level of cross-correlation noise in $[0,2T]$ can allow for a statistically significant transfer entropy, so long as we are beyond a threshold of autocorrelation noise $\alpha\approx T/3$. However, even in this highly sparse regime with little available signal, the reduced transfer entropy is able to detect a regime in which it is not worth transmitting $\bm{y}^{(+1)}$ from $\bm{X}^{(-l)}$. In this case, increasing the number of timesteps $T$ substantially will cause the results to mimic those in the left-hand plot, but for $T=1000$ the autocorrelation in $\bm{y}$ is impossible to detect after only a small amount of noise $\alpha$ is applied.

\setlength{\textfloatsep}{0pt}

\clearpage
\section{Inference of sparse networks with reduced transfer entropy}

In this supplement we show experimental results demonstrating the ability for the proposed reduced transfer entropy measure to infer sparse network structures from coupled time series dynamics.

As a classic example system, we consider a voter model \cite{sood2005voter,schneider2009generalized} taking place on a graph $G=(V,E)$, where each node in $V$ is an agent and directed edges $E$ couple agents that directly influence each other's votes. At each time step $t=1,...,T$, there is a state vector $\bm{s}_t\in \{1,...,C\}^{\abs{V}}$ that stores the vote $s_{ti}\in \{1,...,C\}$ of each agent $i\in V$, and this state vector can update over time as the agents are exposed to their neighbors' opinions. Specifically, an edge $(i,j)\in E$ indicates that agent $i$ can directly influence agent $j$'s vote in each state update, and we denote the set of edges coming into node $i$---i.e., agents that influence $i$---as its in-neighborhood $\partial_i$. The dynamics evolve such that at each successive timestep $t+1$, the agent $i$ adopts a random vote with probability $\epsilon\in [0,1]$ and copies the vote $s_{tj}$ of a random neighbor $j\in \partial_i$ from the previous timestep $t$ with probability $1-\epsilon$. The agents are initially assigned their votes uniformly at random from $\{1,...,C\}$.

We use this voter model to examine the extent to which the proposed transfer entropy method can recover meaningful sparse network structures in known dynamics. For each experiment, we initialize a node set $V$ with $\abs{V}=20$ nodes and generate a synthetic sparse ground truth graph $G_0$ by choosing a fixed number $k\in \{2,5\}$ of nodes in $V$ uniformly at random to form each node $i$'s in-neighborhood $\partial_i$. We then run the voter model described above on $G_0$ to generate a set of time series $\bm{F}^{\abs{V}\times T}$, and aim to recover $G_0$ given only the observed dynamics $\bm{F}$ by applying the different transfer entropy formulations used in Figure 4. We also include the multivariate transfer entropy of Eq.~\ref{eq:MVTE}, greedily adding neighbor time series to $\partial_i$ for each node $i$ while the multivariate transfer entropy exceeds $\mathcal{M}=0$. At this point, adding additional time series to $\partial_i$ is no longer providing a complementary set of predictors for the time series $\bm{x}_i$ at $i$. We set $l=1$ for the transfer entropies since copying neighbors' states will induce cross-correlation at a lag period of $l=1$. 

We quantify the reconstruction accuracy with the graph similarity measure $\text{NMI}(G_{inf},G_0)$ defined in \cite{felippe2024network}, where $\text{NMI}=1$ implies perfect reconstruction of the planted network structure from the dynamics and $\text{NMI}=0$ means the inferred network structure $G_{inf}$ is entirely uncorrelated with the planted structure $G_0$. We vary the level of noise $\epsilon\in [0,1]$ across the simulations to determine how robust the different transfer entropy formulations are for inferring the planted network $G_0$ when different amounts of noise are added to the vote updates. We set $T=1000$ and $C=2$ for the voter model in the simulations.

\begin{figure}[h!]
    \centering
    \includegraphics[width=1\linewidth]{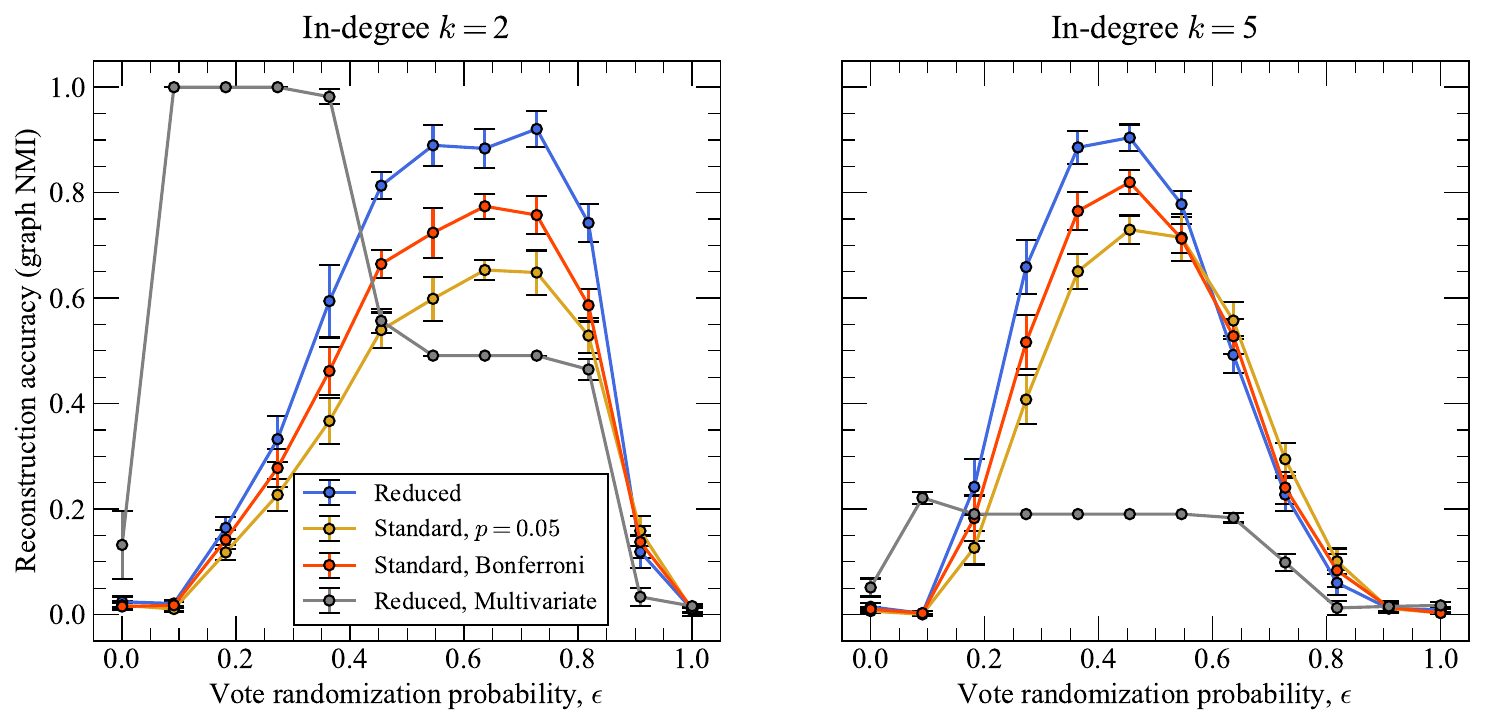}
    \caption{Reconstruction accuracy (graph NMI \cite{felippe2024network}) between planted network structures and those inferred by applying different transfer entropy measures to the observed voter model dynamics, as a function of the vote noise $\epsilon\in [0,1]$. Simulations for random directed networks with fixed in-degrees $k=2$ (left) and $k=5$ (right) are shown.}
    \label{fig:network-reconstruction}
\end{figure}

We show the results of these experiments in Fig.~\ref{fig:network-reconstruction}. In the left panel, we plot the graph NMI versus the vote noise $\epsilon$ for the three methods used in Fig 4 in the main text, as well as for the multivariate transfer entropy of Eq.~\ref{eq:MVTE}. Each data point represents the average of results over $10$ independent simulations of the voter model and network $G_0$, and error bars are two standard errors in the mean. We can see that the original bivariate reduced transfer entropy performs almost strictly better than the standard transfer entropy with $p=0.05$ with and without Bonferroni correction, at all noise levels. Interestingly, these three measures tend to peak in reconstruction accuracy at $\epsilon\approx 0.6$---this is because for low levels of noise, the correlation among the time series decays very slowly across the network, and so each inferred neighborhood $\partial_i$ tends to be fairly large for all methods. On the other hand, we find that the multivariate reduced transfer entropy peaks at low levels of noise $\epsilon>0$ and decreases for larger noise levels (with the exception of $\epsilon=0$, where the whole network has high cross-correlation with $\bm{x}_i$ which also has a high level of autocorrelation). This is because the multivariate transfer entropy only selects time series for $\partial_i$ that provide non-redundant information about $\bm{x}_i$. Since $\bm{x}_i$ copies one neighbor at a time, its neighbors independently provide new information about the future values of $\bm{x}_i$, and so the only neighbors selected by the multivariate transfer entropy tend to be those planted in $\partial_i$. For the right panel in Fig.~\ref{fig:network-reconstruction}, we see a similar trend for the first three methods, with a slight reduction in the value of $\epsilon$ at which the accuracy peaks. However, in this case, the reconstruction accuracy of the multivariate transfer entropy plummets. This is because of the limited number of observations relative to the large embedding dimension provided by additional neighbors---each additional neighbor increases the embedding dimension of the conditioned time series by amultiplicative factor of $C=2$. For this reason, the multivariate transfer entropy tends to return fewer than $k=5$ neighbors in each simulation, as each is very costly from an information perspective. However, increasing the length of the simulations $T$ should improve the performance of the multivariate measure since the total information cost of specifying each time series increases, so including additional neighbors becomes comparatively more affordable.

\clearpage
\section{Selection of optimal lags using reduced transfer entropy}

Here we provide experiments demonstrating the ability for the proposed method to select the optimal lag for two time series according to the Minimum Description Length (MDL) principle.

As the reduced transfer entropy is derived using the MDL principle, we can use this same principle to select the optimal lags $k,l$ for any given observed pair of time series $\bm{x},\bm{y}$. The pair of lags $k,l$ that produces the maximum value of the reduced transfer entropy is the lag configuration at which we achieve the greatest reduction in the description length by using the time series $\bm{x}$ to predict $\bm{y}$. Thus, such a lag configuration is optimal according to the MDL principle.

Here we run experiments similar to those in Figure 2 in the main text, except for each simulation we use the parameters $\{T,l_0,C\}=\{100,5,2\}$ and fix the level of cross-correlation noise to $\beta=T/10$ so that each time series $\bm{x}$ is highly correlated with its target time series $\bm{y}$. We then fix a level of autocorrelation noise $\alpha>\beta$ so that there is an easily detectable non-negative transfer entropy at the planted lag of $l_0=5$, and vary the lag $l$ used for our computation of the transfer entropy computation over the range $l\in [1,10]$.

\begin{figure}[h!]
    \centering
    \includegraphics[width=0.7\linewidth]{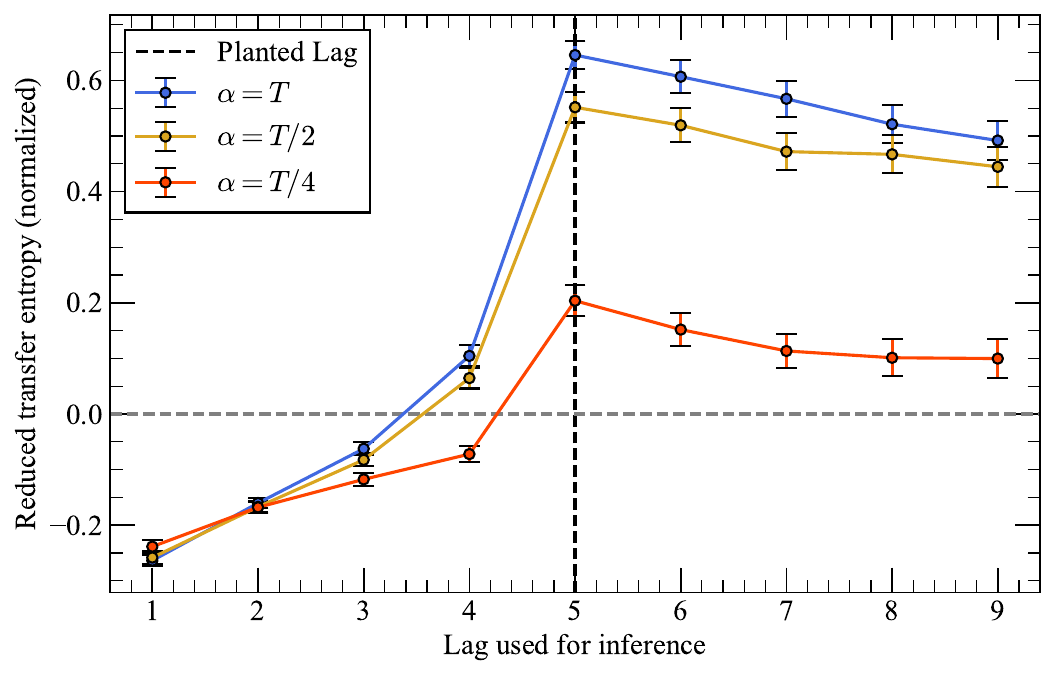}
    \caption{Normalized reduced transfer entropy versus lag $l\in [1,10]$, for synthetic time series generated with planted lag $l_0=5$ and $T,C=100,2$. Each curve represents a different level of planted autocorrelation $\alpha$, and the cross-correlation was set to $\beta=T/10 < \alpha$ for all experiments.}
    \label{fig:lag-selection}
\end{figure}

The results of these experiments are shown in Fig.~\ref{fig:lag-selection}, where we plot the normalized reduced transfer entropy versus the lag $l$ for different values of the autocorrelation $\alpha\in \{T,T/2,T/4\}$. Data points are averages over $1000$ simulations at each value of $l$ and error bars indicate two standard errors in the mean. We can see that for low lags $l<l_0$, the reduced transfer entropy values tend to be negative or near zero, indicating that the time series $\bm{x}$ does not provide a statistically significant amount of additional information about the target series $\bm{y}$. Then, for $l=l_0$, we see the peak reduced transfer entropy values for each $\alpha$, with a decay in the transfer entropy for $l>l_0$. These experiments demonstrate that, under different noise levels, the reduced transfer entropy can robustly identify the optimal lag between two time series.

\clearpage
\section{Transfer entropy network statistics and additional real-world examples}

In this supplement we include experimental results from applying the transfer entropy measures to additional real-world time series datasets. We also provide summary statistics for all three networks studied, in Table I.

We collect the closing price for all stocks in the S\&P 500 over the 10-year period from June 17, 2015 to June 17, 2025, discretizing the data by taking the sign of consecutive daily differences in price to bin values into $\{-1,1\}$. We remove all stocks without data for the entire 10-year period, resulting in $468$ stocks in the final dataset. We then construct transfer entropy networks from these time series using the same procedures as for the air pollution network, setting $p=0.05$ for the permutation testing with the standard transfer entropy and adjusting the Bonferroni correction appropriately for the number of comparisons. 

We also collect armed conflict data from the Armed Conflict Location and Event Data Project \cite{raleigh2010introducing}, which spans the years from 1997-2024. Per the methodology in \cite{kushwaha2023discovering}, we bin the time period into $64$ day windows and assign binary time series values $x_t=1$ for a particular area if there was an armed conflict event during period $t$ in that area, and $x_t=0$ if not. We focus on Nigeria for simpler visualization as done in [6], and utilize the $37$ official Nigerian states ($36+1$ Federal Capital Territory) for administrative area nodes on the networks. We then construct three transfer entropy networks using the same methodology as before. We note that no missing data imputation was needed for any of the transfer entropy network analyses, and that there was negligible variation in the standard transfer entropy network statistics over multiple realizations of the 1000 trials for all networks studied. 

\begin{table}[h]
\centering
\begin{tabular}{|c|c|c|c|c|c|c|}
\hline
Network & $N_{total}$ & $N_{GC}$ & $E_{total}$ & $E_{GC}$ & $\langle k \rangle $ & $\sigma/\expec{k}$ \\
\hline
  AQHI (Reduced TE)       &  18      &    17    &    20    &   20     &   2.2     &   1.6     \\
\hline
  AQHI (Standard TE, $p=0.05$)    &   18     &    18    &    93    &    93    &   10.3     &    2.9    \\
\hline
   AQHI (Standard TE, Bonferroni)      &   18     &    18    &   32     &   32     &    3.6    &    1.5    \\
\hline
\hline
  Stocks (Reduced TE)       &   468     &   459     &   1484     &   1484     &   6.3     &   0.9     \\
\hline
  Stocks (Standard TE, $p=0.05$)    &   468     &   468     &   11908    &    11908    &   50.9     &   2.1     \\
\hline
   Stocks (Standard TE, Bonferroni)      &   468     &  165      &   224     &   187     &   1.0     &     0.4   \\
\hline
\hline
  Conflicts (Reduced TE)       &   37     &    23    &   91     &   91     &   4.9     &    1.2    \\
\hline
  Conflicts (Standard TE, $p=0.05$)    &    37    &   22     &    56     &   56     &   3.0    &    1.1    \\
\hline
   Conflicts (Standard TE, Bonferroni)      &   37     &   2     &    2    &    1    &  0.4      &    $\infty$    \\
\hline
\end{tabular}
\caption{Summary statistics for empirical transfer entropy networks. The recorded statistics are the total number of nodes $N_{total}$; number of nodes in the giant weakly connected component $N_{GC}$; total number of edges $E_{total}$; number of edges in the giant weakly connected component $E_{GC}$; average total (in+out) degree $\expec{k}$; and coefficient of variation in total degrees.}
\label{tab:my_table}
\end{table}

\begin{figure}[h!]
    \centering
    \includegraphics[width=0.36\linewidth]{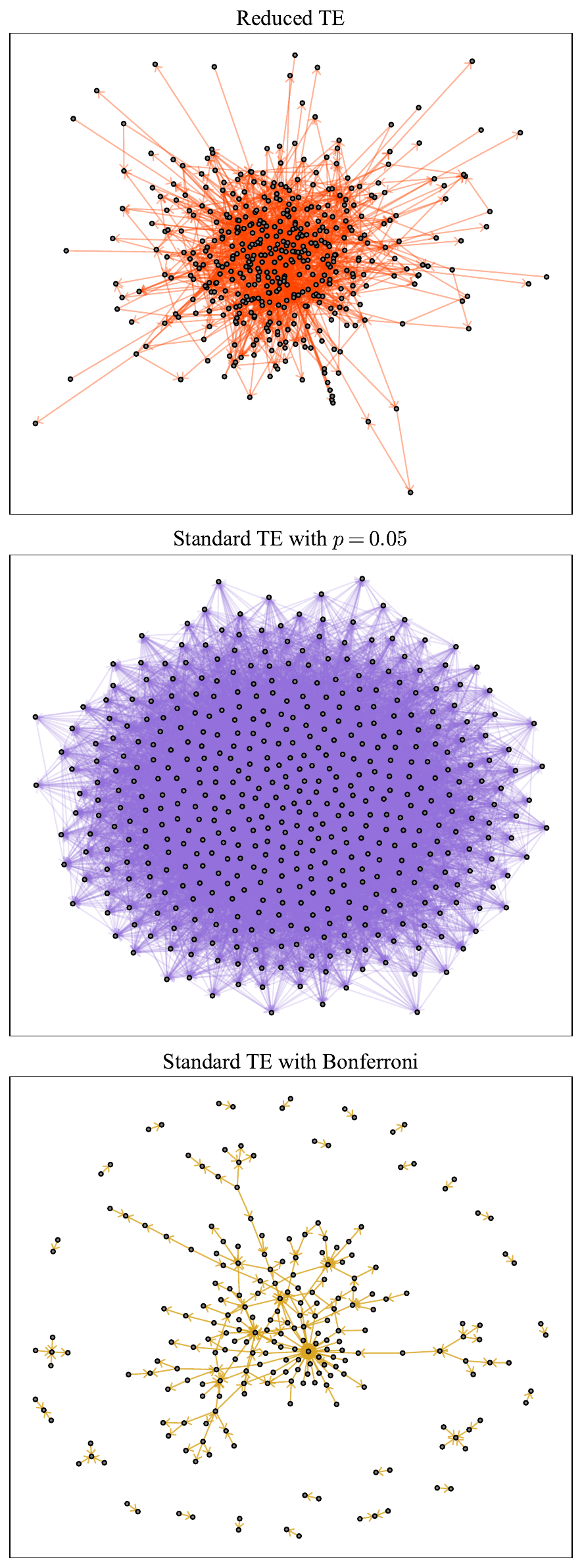}
    \hspace{50pt}
    \includegraphics[width=0.366\linewidth]{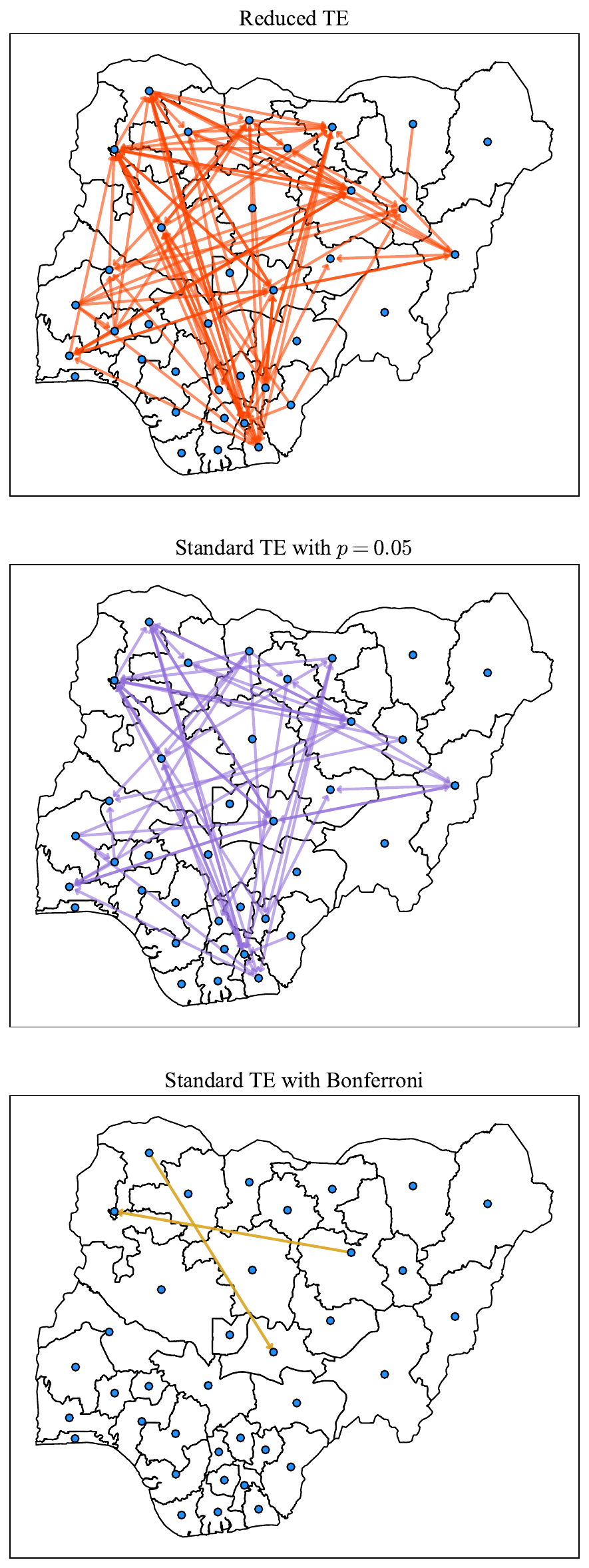}
    \caption{Left: Transfer entropy networks for the S\&P 500 dataset, only retaining nodes with non-zero in- or out-degree for clearer visualization. The reduced transfer entropy network (top) exhibits degree heterogeneity and a core-periphery-type structure. Meanwhile, the standard transfer entropy network with $p=0.05$ (middle) resembles a very dense random directed graph, and a tree-like network with more than half of the nodes isolated when Bonferroni corrected (bottom). Right: Transfer entropy networks for armed conflict events in Nigeria. This time, all nodes are included and placed spatially as centroids of each state, and the reduced transfer entropy gives the densest graph. Summary statistics for these networks and the AQHI network of the main text are shown in Table I.}
    \label{fig:other-real}
\end{figure}

Results for both the stock and armed conflict datasets are shown in Fig.~\ref{fig:other-real}. We can see that in both cases, the reduced transfer entropy measure again gives a useful representation for further analyses, with moderate average degrees and degree heterogeneity as well as a giant component that occupies most of the network. Meanwhile, the standard transfer entropy measure with a $p=0.05$ permutation test significance level produces a very dense graph for the S\&P 500 dataset and a much sparser graph for the armed conflict dataset. The Bonferroni corrected networks are extremely sparse for both cases, indicating that perhaps a less conservative multiple comparisons correction approach is warranted in order to identify any large-scale causal network structure. (The reduced measure does not require this choice as it automatically performs model selection using the MDL principle and data compression.) These examples, together with the example in the main text, suggest that the proposed transfer entropy measure can be an effective method for nonparametrically identifying dependencies in real-world time series data.

\end{document}